\documentclass{JFM-FLM_Au}

\usepackage{lipsum}  
\usepackage{siunitx}
\usepackage{xcolor}
    
    \newcommand{\software}[1]{\textsc{#1}} 

\newcommand{\bgrad}{\boldsymbol{\nabla}} 
\newcommand{\lapl}{\nabla^2} 
\newcommand{\divergence}{\bnabla \bcdot} 
\newcommand{\normalvec}{\mathbf{n}} 
\newcommand{\identityMatrix}{\mathbf{I}} 
\newcommand{\averaged}[1]{\bar{#1}} 
\newcommand{\nondim}[1]{\tilde{#1}} 

\newcommand{\contactangle}{\theta} 
\newcommand{\component}{A} 
\newcommand{\componentB}{B} 
\newcommand{\cvap}{c} 
\newcommand{\cvapEq}{c^\mathrm{eq}} 
\newcommand{\cvapInf}{c^\infty} 
\newcommand{\cvapPure}{c^\mathrm{pure}} 
\newcommand{\molarmass}{x} 
\newcommand{\activitycoef}{\gamma} 
\newcommand{\evapflux}{j} 
\newcommand{\totalflux}{J} 
\newcommand{\Dvap}{D^\mathrm{vap}} 
\newcommand{\massfrac}{y} 
\newcommand{\nondimMassFrac}{\xi} 
\newcommand{\massdensity}{\rho} 
\newcommand{\viscosity}{\mu} 
\newcommand{\surfacetension}{\sigma} 
\newcommand{\curvature}{\kappa} 
\newcommand{\radius}{R} 
\newcommand{\diffusivity}{D} 
\newcommand{\vel}{\mathbf{u}} 
\newcommand{\pres}{p} 
\newcommand{\height}{h} 
\newcommand{\flowrate}{\mathbf{Q}} 
\newcommand{\EvapNumber}{\mathrm{Ev}} 
\newcommand{\EvapNumberTot}{\mathrm{Ev}_\mathrm{tot}} 
\newcommand{\MaraNumber}{\mathrm{Ma}} 
\newcommand{\CapillaryNumber}{\mathrm{Ca}} 
\newcommand{\Deff}{D_\mathrm{eff}} 
\newcommand{\neighDist}{b} 

\newcommand{\shieldingFactor}{F} 
\newcommand{\farFieldDist}{L} 
\newcommand{\stagPoint}{x_0} 

\lefttitle{P. J. Dekker, D. Rocha, D. Lohse, C. Diddens}
\righttitle{Internal flow and concentration in neighbouring evaporating binary droplets and rivulets}

\title{Internal flow and concentration in neighbouring evaporating binary droplets and rivulets}

\author{Pim J. Dekker\aff{1}, Duarte Rocha\aff{1}, Detlef Lohse\aff{1,2}, Christian Diddens\aff{1}}

\affiliation{\aff{1}Physics of Fluids Department, Max-Planck Center Twente for Complex Fluid Dynamics and J. M. Burgers Centre for Fluid Dynamics, University of Twente, 7500AE Enschede, The Netherlands,
\aff{2}Max-Planck Institute for Dynamics and Self-Organization, Am Faßberg 17, 37077 Göttingen, Germany}

\corresau{Pim J. Dekker, \email{p.j.dekker@utwente.nl}, Duarte Rocha, \email{d.rocha@utwente.nl}}

\begin{document}
\maketitle


\begin{abstract}
In this paper, the evaporation of neighbouring multi-component droplets or rivulets -- often found in applications such as inkjet printing, spray cooling, and pesticide delivery -- is studied numerically and theoretically.
The proximity induces a shielding effect that reduces individual evaporation rates and disrupts the symmetry of both the concentration profile and the flow field in the liquids. 
We examine how the symmetry of flow and concentration fields is affected by key parameters, namely the contact angle, the inter-droplet (or inter-rivulet) distance, and the magnitude of surface tension gradient forces (i.e. the Marangoni number). 
We focus on binary mixtures, such as water and 1,2-hexanediol, where only one component evaporates and evaporation is slow, thereby allowing simplifications to the governing equations. 
To manage the complexity of the full three-dimensional droplet problem, we begin with a two-dimensional model of neighbouring rivulets. 
Solving the complete transient equations for rivulets with pinned contact lines and fixed inter-rivulet distance reveals that the asymmetry -- quantified by the position of the interfacial stagnation point of the flow -- diminishes over time. 
Using a validated quasi-stationary model, we find, with increasing contact angle and inter-rivulet distance, that the stagnation point migrates closer to the centre, yet it remains unaffected by the Marangoni number. 
A simplified lubrication model applied to droplets shows similar dependencies on contact angle and distance, although here the stagnation point appears to vary with the Marangoni number. 
We attribute this dependence to the additional azimuthal flow in droplets, leading to a non-linear evolution of the concentration and therefore a non-trivial dependence of the symmetry on the Marangoni number.

\end{abstract}



\clearpage


\section{Introduction}
Applications such as inkjet printing, spray cooling and pesticide delivery often involve the evaporation of neighbouring multi-component droplets or rivulets. 
In these situations, the proximity of the liquid bodies influences the evaporation rate of each droplet or rivulet, thereby altering the flow dynamics within the liquid phase \citep{schofield2020shielding, wray2020competitive, masoud2021evaporation}. 
A natural first step in understanding these complex systems is to study the evaporation of a single liquid body -- a problem that has been extensively investigated for droplets \citep{sefiane2014patterns, brutin2018recent, zang2019evaporation, lohse2020physicochemical, gelderblom2022evaporation, wilson2023evaporation} as well as for rivulets \citep{yarin2006lines, schofield2020shielding}.

For evaporating pure sessile droplets (and rivulets) with a pinned contact line, the geometric singularity at the contact line gives rise to the ``coffee-stain'' effect \citep{deegan1997capillary, deegan2000contact, popov2005evaporative}, where liquid and any suspended particles are drawn towards the pinned contact line to compensate for a higher evaporative loss in that region, resulting in a coffee-ring-like deposit. 
This is often undesirable in applications such as inkjet printing \citep{lohse2022fundamental}. 
In the case of a multi-component droplet (or rivulet), selective evaporation can occur. 
This selective evaporation induces surface tension gradients, which in turn generate Marangoni stresses at the liquid-gas interface \citep{scriven1960marangoni, hu2005analysis, diddens2017evaporating, kim2018direct, marin2019solutal}. 
The resulting flow generated within the liquid is often much stronger than the flow caused by the ``coffee-stain'' effect and can either enhance or suppress it, depending on the direction of the Marangoni flow \citep{hu2006marangoni, diddens2017evaporating, mampallil2018review}.

In real-world applications, however, the liquid bodies are rarely isolated and are often found in proximity. 
In such cases, the vapour concentration in the gas phase surrounding the evaporating components is increased near the liquid bodies.
This reduces the evaporation rate -- characterising the so-called shielding effect -- and extends the lifetime of the liquids \citep{fabrikant1985potential, laghezza2016collective, khilifi2019study, hatte2019universal, chong2020convection, schofield2020shielding, masoud2021evaporation}. This shielding effect disrupts the symmetry of the evaporation-driven flows \citep{wray2020competitive, masoud2021evaporation}.

While capillary flows in neighbouring pure droplets have been widely studied, Marangoni flows induced by the presence of a second component in the liquid has received less attention. 
\citet{cira2015vapour} observed that neighbouring droplets may either attract -- under conditions that promote Marangoni contraction \citep{karpitschka2017marangoni} -- or repel one another, depending on the volatility and composition. 
Moreover, \citet{dekker2024pinning} reported that preferential pinning on the side further from an adjacent droplet can cause the centres of the droplets to move apart.

Our work builds on the experiments of \citet{dekker2024pinning}, in which the evaporation of two neighbouring droplets of a binary mixture of water and 1,2-hexanediol was investigated. 
In particular, we develop numerical simulations for both two evaporating droplets and their two-dimensional counterpart, two evaporating rivulets, in order to identify the key parameters that govern the symmetry of the flow field. 
By employing justified simplifications, we demonstrate that the asymmetry in the internal flow and concentration fields is primarily controlled by three parameters: the contact angle $\contactangle$, the inter-droplet (or inter-rivulet) distance $b$, and the Marangoni number $\MaraNumber$. 
Notably, for rivulets, the asymmetry is nearly independent on the Marangoni number, whereas for droplets the dependence is more marked.

This paper is structured as follows:
In \S~\ref{sec:governing_equations}, we introduce the governing equations for the evaporation of rivulets and droplets, and we present transient numerical simulations for rivulets.
The slow evaporation motivates the introduction of a simplified quasi-stationary model (\S~\ref{sec:qs_model}) and its lubrication version (\S~\ref{sec:lubrication}). 
Section~\ref{sec:evaporative_flux} examines the local evaporative flux for droplets (\S~\ref{sec:evaporative_flux_droplet}) and rivulets (\S~\ref{sec:evaporative_flux_rivulet}), to be used in the simplified models. 
We then validate the models against transient simulations (\S~\ref{sec:validation}) and investigate the stagnation point shift, for rivulets (\S~\ref{sec:shift_stagnation_point_rivulet}, with some analytical results in \S~\ref{sec:analytical_rivulet}) and droplets (\S~\ref{sec:shift_stagnation_point_droplet}).
Finally, we conclude in \S~\ref{sec:conclusions}.


\section{Transient system of equations for evaporating rivulets and droplets} \label{sec:transient_model}

\subsection{Underlying dynamical equations} \label{sec:governing_equations}

We focus on the evaporation of a single or two neighbouring binary droplets (or rivulets) which form a low contact angle $\contactangle$ with the substrate. 
Let us consider that: only one component $\componentB$ in the liquid phase can evaporate into the vapour phase; the surface tension monotonically decreases with the concentration of the non-volatile component; and evaporation is slow.
Under these assumptions, we avoid complexities arising from Marangoni instabilities \citep{diddens2017evaporating, diddens2024non, rocha2024marangoni} and we meet the exact conditions to apply the models introduced by \citet{diddens2021competing}, which we describe in the following.

At ambient conditions, the evaporation is dominated by the diffusion of vapour in the surrounding air. 
The solution of the Laplace equation
\begin{equation}\label{eq:laplace_cvap}
  \lapl \cvap_\componentB = 0,
\end{equation}
with the boundary conditions
\begin{equation}
  \cvap_\componentB = \cvapEq_\componentB = \cvapPure_\componentB \activitycoef_\componentB \molarmass_\componentB \quad \text{at liquid-gas interface},
\end{equation}
\begin{equation}
  \cvap_\componentB = \cvapInf_\componentB \quad \text{at far-field},
\end{equation}
gives the vapour concentration $\cvap_\componentB$ in the gas phase.  Here, $\cvapInf_\componentB$ is the vapour concentration at far-field, dependent of the relative humidity, $\activitycoef_\componentB$ is the activity coefficient, and $\molarmass_\componentB$ is the molar mass of component $\componentB$.

The diffusive mass flux of the component $\componentB$ through the liquid-gas interface due to evaporation is given by
\begin{equation}
  \evapflux_\componentB = -\Dvap_\componentB \bgrad \cvap_\componentB \bcdot \normalvec ,
\end{equation}
where $\Dvap_\componentB$ is the diffusion coefficient of vapour and $\normalvec$ is the normal vector at the liquid-gas interface. If only one component evaporates, then the total evaporation rate is $\evapflux = \evapflux_\componentB$. Naturally, the mass fraction $\massfrac_\component$ of the non-volatile component $\component$ in the liquid phase increases during evaporation, which is accounted for by the Robin boundary condition in the co-moving frame of the liquid-gas interface
\begin{equation}
  \massdensity \diffusivity \bgrad \massfrac_\component \bcdot \normalvec = \massfrac_\component \evapflux ,
  \label{eq:robinbc}
\end{equation}
where $\massdensity$ is the mass density of the liquid phase and $\diffusivity$ is the diffusivity in the liquid phase. Within the liquid, the mass fraction $\massfrac_\component$ is governed by the advection-diffusion equation
\begin{equation}
  \rho\left(\partial_t \massfrac_\component + \vel \bcdot \bgrad \massfrac_\component\right) = \bgrad \cdot (\rho D \bgrad \massfrac_\component).
\end{equation}
Navier-Stokes flow without body forces in the liquid phase is described by
\begin{equation}
  \partial_t \massdensity + \bgrad \bcdot (\massdensity \vel) = 0,
\end{equation}
\begin{equation}
  \massdensity (\partial_t \vel + \vel \bcdot \bgrad \vel)= - \bgrad \pres + \bnabla \bcdot \left( \viscosity \bnabla \vel\right),
\end{equation}
where $\viscosity$ is the viscosity of the liquid phase and $\pres$ is the pressure. We neglect gravity since the droplet is small and the contact angle is low. At the substrate, the no-slip boundary condition is applied, i.e. we consider a constant contact radius \citep{picknett1977evaporation}. 
At the liquid-gas interface, the kinematic boundary condition
\begin{equation}
  \massdensity (\vel - \vel_I) \bcdot \normalvec = \evapflux,
\end{equation}
and the dynamic boundary condition
\begin{equation}\label{eq:dynamic_bc}
  \normalvec \bcdot \left( \viscosity (\bgrad \vel + \bgrad \vel^T) - \pres \identityMatrix \right) = \normalvec \surfacetension \curvature + \bgrad_t \surfacetension,
\end{equation}
are applied, where $\vel_I$ is the velocity of the liquid-gas interface, $\surfacetension$ is the surface tension, $\curvature$ is the curvature of the liquid-gas interface and $\bgrad_t$ is the surface gradient operator. 
In this manuscript, we neglect the influence of temperature on flow field, for simplicity. 
All properties are therefore only dependent on the mass fraction $\massfrac_\component$.


\subsection{Numerical simulations results for rivulets} \label{sec:transient_rivulet}

Due to the non-axisymmetric evaporation rate of neighbouring droplets, the resulting flow field is asymmetric. 
Accurately computing the flow in this scenario requires solving the full three-dimensional problem, which is computationally expensive. 
As a starting point, we solve the full two-dimensional problem for rivulets. 
In later sections, we introduce simplifications to develop a model for droplets. 
All differential operators introduced in the previous section are expressed in Cartesian coordinates for the rivulet geometry. 
Note that the solution for $\cvap$ in the gas phase within a two-dimensional domain depends on the distance $\farFieldDist$ to the far-field boundary \citep{sneddon1966mixed, masoud2009analytical, schofield2020shielding}. 
We fix the far-field boundary at a distance $\farFieldDist / \radius = 100$.

\begin{figure}[ht!]
\centering
\includegraphics[width=1\textwidth]{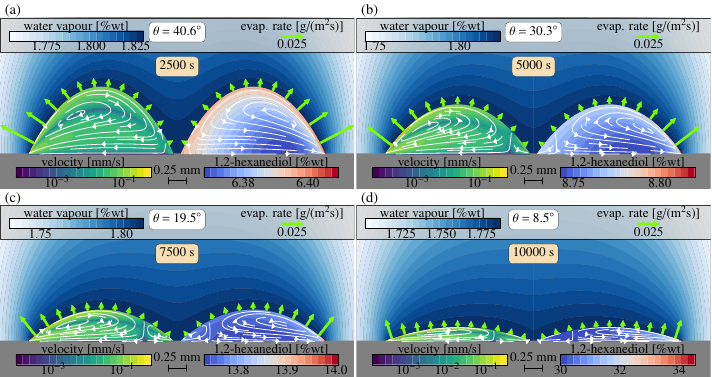}
\caption{Transient direct numerical simulations, evaluated after \SI{2500}{\second} (a), \SI{5000}{\second} (b), \SI{7500}{\second} (c), and \SI{10000}{\second} (d), of two neighbouring rivulets, with an initial contact angle of $\contactangle = \SI{50}{\degree}$, a 1,2-hexanediol mass fraction of $\massfrac_\componentB = \SI{5}{\percent}$ and a distance $b=$\SI{2.1}{\milli\meter} between their centres, evaporating in an environment with relative humidity of \SI{70}{\percent}.
The plots are magnified vertically by 2.5 $\times$, in order to better visualise the flow inside the rivulet.
On the left side of each subfigure, the velocity magnitude is displayed with superimposed streamlines.
On the right, the 1,2-hexanediol mass fraction is shown. 
Additionally, the gas phase illustrates the water vapour mass fraction, while the liquid-gas interface highlights the evaporation flux of water.
The $x$-position where the two vortices in the rivulet meet (which we refer to as the stagnation point $\stagPoint$) increases with time.
}
\label{fig:transient_rivulet}
\end{figure}

The system of equations described in \S~\ref{sec:transient_model} is solved using a sharp-interface Lagrangian--Eulerian finite element method implemented in the software package \software{pyoomph} \citep{diddens2024bifurcation}, which is based on \software{oomph-lib} \citep{heil2006oomph} and \software{GiNaC} \citep{bauer2002introduction}. 
All computational domains are discretised using meshes composed of triangular elements. Linear basis functions are employed for pressure $\pres$ fields, whereas quadratic basis functions are used for the velocity $\vel$, vapour concentration $\cvap$ and mass fraction $\massfrac_\component$ fields.
Time integration is performed using a second-order implicit backward differentiation formula.

At early stages of evaporation, the streamlines in the rivulet nearly form a single vortex (see Figure~\ref{fig:transient_rivulet}). 
We define the stagnation point, $\stagPoint$, as the x-coordinate where the two vortices merge -- identified by a zero value in the tangential interfacial velocity, $\vel_t = \vel \bcdot \mathbf{t}$. 
For flat rivulets, this zero-velocity condition is nearly uniform along the vertical direction, which justifies using $\vel_t$ to determine $\stagPoint$. 
Alternatively, we could also find $\stagPoint$ by identifying the point where the velocity close to the substrate crosses zero, but, at least for rivulets, this would yield virtually the same result, particularly for low contact angles.
As evaporation progresses, the stagnation point shifts toward the centre of the rivulet, prompting an investigation into how $\stagPoint$ depends on the contact angle, surface tension forces, and the distance between the droplets.
We begin by introducing a simplified quasi-stationary system of equations, allowing us to identify the relevant dimensionless control parameters.


\section{Simplified quasi-stationary system of equations for evaporating rivulets and droplets}\label{sec:simplified_model}


\subsection{Quasi-stationary model}\label{sec:qs_model}

We focus on the evaporation of droplets (or rivulets) with quasi-stationary evaporation, such as mixtures of water and 1,2-hexanediol \footnote{We ignore the slight nonmonotonicity of the surface tension as a function of the 1,2-hexanediol \citep{diddens2024non}. Though it leads to extremely rich interesting behaviour \citep{diddens2024non}, here it distracts from the main point of this paper. So our binary liquid can be seen as a model liquid with material parameters inspired by 1,2-hexanediol-water mixtures.}.
In such cases, the internal flow within the liquid remains stable and approximately quasi-steady at each instant during the drying process \citep{diddens2021competing}, provided that the solutal Marangoni flow induces sufficient mixing. 
Consequently, we assume small compositional deviations around the spatially averaged mass fraction $\massfrac_{\component,0}$, and express the local mass fraction of component $\component$ as $\massfrac_\component = \massfrac_{\component,0} + \massfrac$, where $\massfrac$ denotes the deviation from the mean.
We also consider, for simplicity, that changes in temperature do not significantly affect the flow field when compared to the effects of composition.

All liquid properties are considered constant, evaluated at the spatially averaged composition and ambient temperature, except for the liquid-gas surface tension, which varies according to the first order Taylor expansion: 
$\surfacetension(\massfrac_\component) = \surfacetension(\massfrac_{\component,0}) + \massfrac \, \partial_{\massfrac_\component} \surfacetension$. 
Since the average composition evolves slowly -- at least for droplets or rivulets with contact angles $\contactangle \gtrsim \SI{5}{\degree}$ -- this expansion can be performed at any given time during the drying process. We introduce the following nondimensional scales:
\begin{equation*}
  x = \radius \nondim{x}, \quad 
  t = \frac{\radius^2}{\diffusivity_0} \nondim{t}, \quad 
  \vel = \frac{\diffusivity_0}{\radius} \nondim{\vel}, \quad 
  p = \frac{\viscosity_0 \diffusivity_0}{\radius^2} \nondim{\pres},
\end{equation*}
where $\diffusivity_0$ and $\viscosity_0$ are the diffusivity and viscosity of the liquid, respectively, and $\radius$ is the base radius of the droplet or rivulet. These nondimensional scales are, in principle, time-dependent, as the liquid volume $V$ and the average composition $\massfrac_{\component,0}$ evolve. However, as shown by \citep{diddens2021competing}, the dynamics can be considered quasi-stationary at each instant during evaporation, at least after the initial short transient and as long as Marangoni flow is strong enough.

The vapour concentration $\cvap_\componentB$ in the gas phase is then expressed as:
\begin{equation}
  \cvap_\componentB = (\cvapEq_{\componentB,0} - \cvapInf_\componentB) \nondim{\cvap} + \cvapInf_\componentB.
\end{equation}
Note that we neglect the effect of the variations on the local composition on the vapour concentration. 
As argued by \citet{diddens2021competing}, if the local composition at the interface is close to the average composition, which happens in droplets or rivulets with strong Marangoni induced circulation, then the equilibrium vapour concentration at the interface is approximately constant. 
Furthermore, for the particular case of 1,2-hexanediol and water mixtures, the equilibrium vapour concentration remains roughly constant for 1,2-hexanediol mass fractions below 0.5 \citep{dekker2024pinning}, based on UNIFAC activity coefficients \citep{constantinescu2016further}.
The Laplace equation~\eqref{eq:laplace_cvap} is then given by 
\begin{equation}\label{eq:nondim_laplace_cvap}
  \nondim{\lapl} \nondim{\cvap} = 0, 
\end{equation}
with boundary conditions $\nondim{\cvap} = 1$ at the liquid-gas interface, and $\nondim{\cvap} = 0$ at the far-field. Accordingly, the evaporation flux $\evapflux$ can be expressed as
\begin{equation}
  \evapflux_\componentB = \frac{\Dvap_\componentB}{\radius}  
    (\cvapEq_{\componentB,0} - \cvapInf_\componentB) \nondim{\evapflux},
\end{equation}
where $\nondim{\evapflux} = -\nondim{\partial}_n \nondim{\cvap}$ depends solely on the liquid geometry. Neglecting higher-order terms in $\massfrac$, the advection-diffusion equation for the deviation $\massfrac$ becomes:
\begin{equation}\label{eq:nondim_advection_diffusion}
  \partial_{\nondim{t}} \massfrac_{\component,0} 
  + \partial_{\nondim{t}} \massfrac 
  + \nondim{\vel} \bcdot \nondim{\bgrad} \massfrac 
  = \nondim{\lapl} \massfrac,
\end{equation}
with the boundary condition at the liquid-gas interface:
\begin{equation}\label{eq:nondim_robin_bc}
  - \nondim{\bgrad} y \bcdot \normalvec 
  = \EvapNumber \evapflux - \EvapNumberTot y \evapflux.
\end{equation}
The dimensionless evaporation numbers are defined as:
\begin{gather}
  \EvapNumber = \frac{- y_{A,0} \Dvap_B (\cvapEq_{B,0} - \cvapInf_B)}{\massdensity_0 \diffusivity_0}, \quad \EvapNumberTot = \frac{\Dvap_B (\cvapEq_{B,0} - \cvapInf_B)}{\massdensity_0 \diffusivity_0}.
\end{gather}

The parameter $\EvapNumber$ quantifies the strength of the concentration gradients induced by the selective evaporation of component $\componentB$ and thereby the Marangoni stresses driving the flow. It plays a central role in determining the flow field.

The parameter $\EvapNumberTot$ measures the total evaporation rate and is associated with flow towards the contact line in pinned liquids, i.e.\ the “coffee-stain” effect. This contribution is generally small under strong Marangoni convection and minimal compositional deviation. However, towards the end of the drying process, when the non-volatile component dominates, local compositional effects become more significant. 
For simplicity, following the well-established model of \citet{diddens2021competing}, we assume $\EvapNumberTot = 0$. The kinematic boundary condition at the liquid-gas interface is then reduced to
\begin{equation}\label{eq:nondim_kinbc}
  \normalvec \bcdot \nondim{\vel}  = 0 .
\end{equation}

Introducing the scaled variable $\nondimMassFrac = \frac{y}{\EvapNumber}$, the quasi-stationary form of equation \ref{eq:nondim_advection_diffusion} becomes
\begin{equation}
  \nondim{\vel} \bcdot \nondim{\bgrad} \nondimMassFrac = \nondim{\lapl} \nondimMassFrac + \dfrac{\nondim{\totalflux}}{\nondim{V}},
\end{equation}
where $\nondim{\totalflux}/\nondim{V}$ is the integrated evaporation flux $\nondim{\totalflux}$ over the liquid-gas interface, averaged by the nondimensional volume $\nondim{V}$ (or cross-sectional area for a rivulet).
The Robin boundary condition \eqref{eq:nondim_robin_bc} at the liquid-gas interface becomes the Neumann condition
\begin{equation}\label{eq:nondim_robin_bc2}
  - \nondim{\bgrad} \nondimMassFrac \bcdot \normalvec = \nondim{\evapflux}.
\end{equation}
The dynamic boundary condition \ref{eq:dynamic_bc} at the interface is modified to
\begin{equation}\label{eq:dynamic_bc2}
  \normalvec \bcdot \left( \nondim{\bgrad} \nondim{\vel} + \nondim{\bgrad} \nondim{\vel}^T \right) = \normalvec \frac{\nondim{\curvature}}{\CapillaryNumber} + \MaraNumber \nondim{\bgrad}_t \nondimMassFrac,
\end{equation}
where the dimensionless capillary number $\CapillaryNumber$ and Marangoni number $\MaraNumber$ are defined as
\begin{equation}
  \CapillaryNumber = \frac{\viscosity_0 \diffusivity_0}{\surfacetension_0 \radius}, \quad 
  \MaraNumber = \frac{R \partial_{\massfrac_\component} \surfacetension}{\viscosity_0 \diffusivity_0} \EvapNumber.
\end{equation}
For small droplets or rivulets, the capillary number is small. In that case, the liquid will remain with a spherical cap.

Since the flow's characteristic Schmidt number $\mathrm{Sc} = \viscosity_0/(\diffusivity_0 \massdensity_0)$ is typically large and the Reynolds number small, we can neglect the inertial terms in the Navier-Stokes equation. We therefore consider incompressible Stokes flow in the liquid phase to be given by
\begin{equation}\label{eq:nondim_incomp}
  \nondim{\divergence} \nondim{\vel} = 0,
\end{equation}
\begin{equation}\label{eq:nondim_stokes}
  \nondim{\bgrad} \nondim{p} = \nondim{\lapl} \nondim{\vel}.
\end{equation}

Naturally, for droplets, the system of equations 
\eqref{eq:nondim_laplace_cvap},\eqref{eq:nondim_kinbc}--\eqref{eq:nondim_stokes} (hereafter termed the quasi-stationary model) is still three-dimensional and therefore computationally expensive. 
However, in the limit of small contact angles, the lubrication approximation, introduced in the following subsection, can be employed to effectively reduce the problem to two dimensions.


\subsection{Lubrication quasi-stationary model} \label{sec:lubrication}

In the lubrication quasi-stationary model (hereafter simply lubrication model), we redefine some nondimensional scales to make use of the small vertical length (as compared to the horizontal one) scale of the liquid body:
\begin{equation*}
  \height = \contactangle \radius \nondim{\height}, \quad \vel_{\Vert} = \frac{\diffusivity_0}{\radius} \nondim{\vel}_{\Vert}, \quad u_z = \contactangle \frac{\diffusivity_0}{\radius} \nondim{u}_z, \quad p = \frac{\viscosity_0 \diffusivity_0}{\contactangle^2 \radius^2} \nondim{\pres},
\end{equation*}
where the subscript $\Vert$ indicates the horizontal component of the vector. For brevity, in the following, we drop the horizontal component subscript in the velocity, in the gradient $\bgrad$ and divergence $\divergence$ operators.

Following the standard derivation of the thin film approximation \citep{eggers2010nonlocal,diddens2017modeling} and assuming negligible ``coffee-stain'' flow ($\EvapNumberTot = 0$), we arrive at the following evolution equation for the liquid height $h$:
\begin{equation}\label{eq:nondim_film_thickness}
  \nondim{\bgrad} \bcdot \nondim{\flowrate} = 0
\end{equation}
where
\begin{equation}\label{eq:nondim_flowrate}
  \nondim{\flowrate} = \int_0^{\nondim{\height}} \nondim{\vel} d\nondim{z} = - \dfrac{\nondim{\height}^3}{3} \nondim{\bgrad} \nondim{\pres} + \MaraNumber \contactangle \dfrac{\nondim{\height}^2}{2} \nondim{\bgrad} \averaged{\nondimMassFrac}.
\end{equation}
We consider the limit of zero-$\CapillaryNumber$, i.e.\ the droplet (or rivulet) remains a spherical cap. This means that the Laplace pressure is sufficiently strong to prevent any notable deformations. In droplets (or rivulets) with small contact angle, strong Marangoni flow can lead to, depending on its direction, a shape contraction \citep{karpitschka2017marangoni} or a pancake-like shape \citep{pahlavan2021evaporation}.

For the compositional evolution, we consider a regime where vertical gradients of composition are small, but not negligible. We consider the evolution equation for the vertically averaged mass fraction $\averaged{\nondimMassFrac}$ which includes Taylor-Aris dispersion \citep{ramirez-soto_taylor_2022}:
\begin{equation}\label{eq:nondim_mass_fraction}
  \nondim{\flowrate} \bcdot \nondim{\bgrad} \averaged{\nondimMassFrac} = \nondim{\bgrad} \bcdot (\Deff \nondim{\height} \nondim{\bgrad} \averaged{\nondimMassFrac}) - \dfrac{\nondim{\evapflux}}{\contactangle} + \dfrac{\nondim{\totalflux}\nondim{\height}}{\nondim{V}},
\end{equation}
where the dimensionless effective diffusivity $\Deff$ is given by
\begin{equation}\label{eq:Deff}
  \Deff = 1 + \contactangle^2 \left( \dfrac{2 \nondim{\flowrate}_C^2}{105} + \dfrac{\nondim{\flowrate}_C \bcdot \nondim{\flowrate}_M}{20} + \dfrac{\nondim{\flowrate}_M^2}{30} \right),
\end{equation}

\begin{equation}
  \text{with } \quad \nondim{\flowrate}_C = -\dfrac{\nondim{\height}^3}{3} \nondim{\bgrad} \nondim{\pres}, \quad \nondim{\flowrate}_M = \contactangle \MaraNumber \dfrac{\nondim{\height}^2}{2} \nondim{\bgrad} \averaged{\nondimMassFrac}.
\end{equation}

Within the lubrication model, the local evaporative flux cannot be computed directly because doing so would require explicitly modelling the vapour concentration $\cvap_\componentB$ in the gas phase. 
Instead, this flux must be imposed as a boundary condition at the liquid-gas interface. 

\section{Local evaporative flux of neighbouring drops and rivulets for the lubrication model}\label{sec:evaporative_flux}

\subsection{Droplets}\label{sec:evaporative_flux_droplet}

For flat pure droplets, the evaporation flux is given by \citep{popov2005evaporative}:
\begin{equation}\label{eq:evapfluxbase_droplet}
  \evapflux_{B,0} = \dfrac{2 \Dvap_\componentB (\cvapEq_{\componentB,0} - \cvapInf_\componentB)}{\pi \radius \sqrt{1 - r^2}}.
\end{equation}
For the sake of simplicity, we consider that the presence of a second component in the droplet does not significantly affect the evaporation flux of the first component. In the particular case of 1,2-hexanediol and water mixtures, the equilibrium vapour concentration remains roughly constant for 1,2-hexanediol mass fractions below 0.5, which justifies the above assumption.
For increased accuracy, we consider the analytical integral solution \citep{popov2005evaporative} for the evaporation flux of the droplet, considering the effect of the contact angle.

When another droplet is present, the evaporation flux is modified due to a shielding effect. The shielding factor $\shieldingFactor$, which multiplied by the base evaporative flux of eq.\ \ref{eq:evapfluxbase_droplet} gives the evaporation rate $\evapflux_{B,1}$ of each neighbouring droplet, can be written as \citep{wray2020competitive}:
\begin{equation}
  \shieldingFactor = 1 - \dfrac{4 \sqrt{\neighDist^2 - 1}}{\left(1 + \frac{2}{\pi} \arcsin{\frac{1}{\neighDist}}\right)\left(2\pi (r^2 + \neighDist^2 - 2r \neighDist \cos{\phi})\right)} ,
\end{equation}
where $\neighDist$ is the distance between the centres of the two droplets and $\phi$ the azimuthal angle in the drop. The shielding factor is independent of $\contactangle$ since it was derived for flat drops.

\subsection{Rivulets}\label{sec:evaporative_flux_rivulet}

\begin{figure}[ht!]
\centering
\includegraphics[width=1\textwidth]{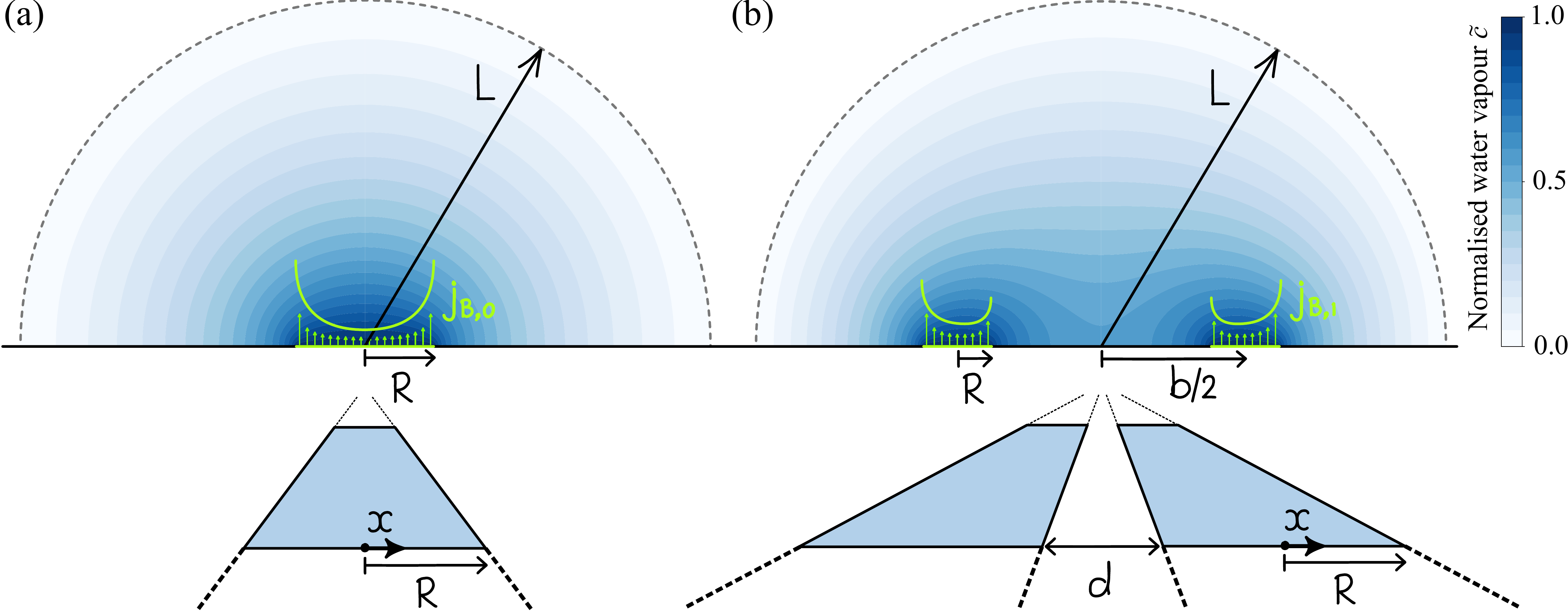}
\caption{Sketch of (a) isolated and (b) neighbouring evaporating rivulets in an infinite half-space of size $L$. In this sketch we have chosen a small value for $L$ for illustration purposes. The top row shows the numerically calculated water vapour concentration in the entire domain. The green arrows indicate the local evaporative flux. 
The bottom row shows a sketch of the rivulet geometry. The origin of the x-coordinate is chosen at the centre of the rivulet and the right rivulet in the neighbouring case. $R$ is the radius, or half-width, of the rivulet. The distance between rivulets can be defined as the centre-to-centre distance $b$ or the edge-to-edge distance $d$.}
\label{fig:flux_sketch}
\end{figure}

Figure \ref{fig:flux_sketch} shows the rivulet geometry, the local evaporative flux, and the water vapour concentration in the semi-circular infinite half-space.
To the best of the authors' knowledge, the evaporation flux of a rivulet for an arbitrary contact angle has not been derived yet.
\citet{schofield2020shielding} showed the non-existence of an analytical solution for the vapour concentration in the gas phase with and infinite half-space, due to the logarithmic nature of the solution of the two-dimensional Laplace equation.
When considering a finite semi-elliptic domain with radii $\farFieldDist$ and points at $\radius$ (approximately a semi-circular domain for large enough $\farFieldDist$), the evaporation flux of a flat isolated rivulet is given by
\begin{equation}\label{eq:evapfluxbase_rivulet}
  \evapflux_{B,0} = \dfrac{\Dvap_\componentB (\cvapEq_{\componentB,0} - \cvapInf_\componentB)}{ \mathrm{arcsinh}{\left(\farFieldDist/\radius\right)}}\frac{1}{\sqrt{\radius^2 - x^2}},
\end{equation}
where $x$ is the distance from the centre of the rivulet to the point of interest. Note that $\mathrm{arcsinh}{\left(\farFieldDist/\radius\right)} \approx \ln{\left(2\farFieldDist/\radius\right)}$ for $\farFieldDist \gg \radius$. For two neighbouring rivulets the evaporative flux is given by

\begin{equation}\label{eq:evapflux_rivulet}
  \evapflux_{B,1} =\frac{\Dvap_\componentB (\cvapEq_{\componentB,0} - \cvapInf_\componentB)}{2\,\mathrm{arcsinh}\left(L/\sqrt{2 b R}\right)}\frac{1}{\sqrt{\radius^2 - x^2}}\frac{b+2x}{\sqrt{(x+b+R)(x+b-R)}}
\end{equation}
Here, $x$ is the distance from the centre of the right rivulet, as indicated in figure \ref{fig:flux_sketch}. Note that \citet{schofield2020shielding} uses the coordinate $\eta$, which is centred between both rivulets.
The first term in eq. \eqref{eq:evapflux_rivulet} is the integral flux, the second term the local flux for an isolated rivulet, and the final term is the local shielding effect.
We consider \eqref{eq:evapflux_rivulet} when applying the lubrication model to neighbouring rivulets.

\section{Validation of different models and approximations} \label{sec:validation}

\begin{figure}[ht!]
\centering
\includegraphics[width=1\textwidth]{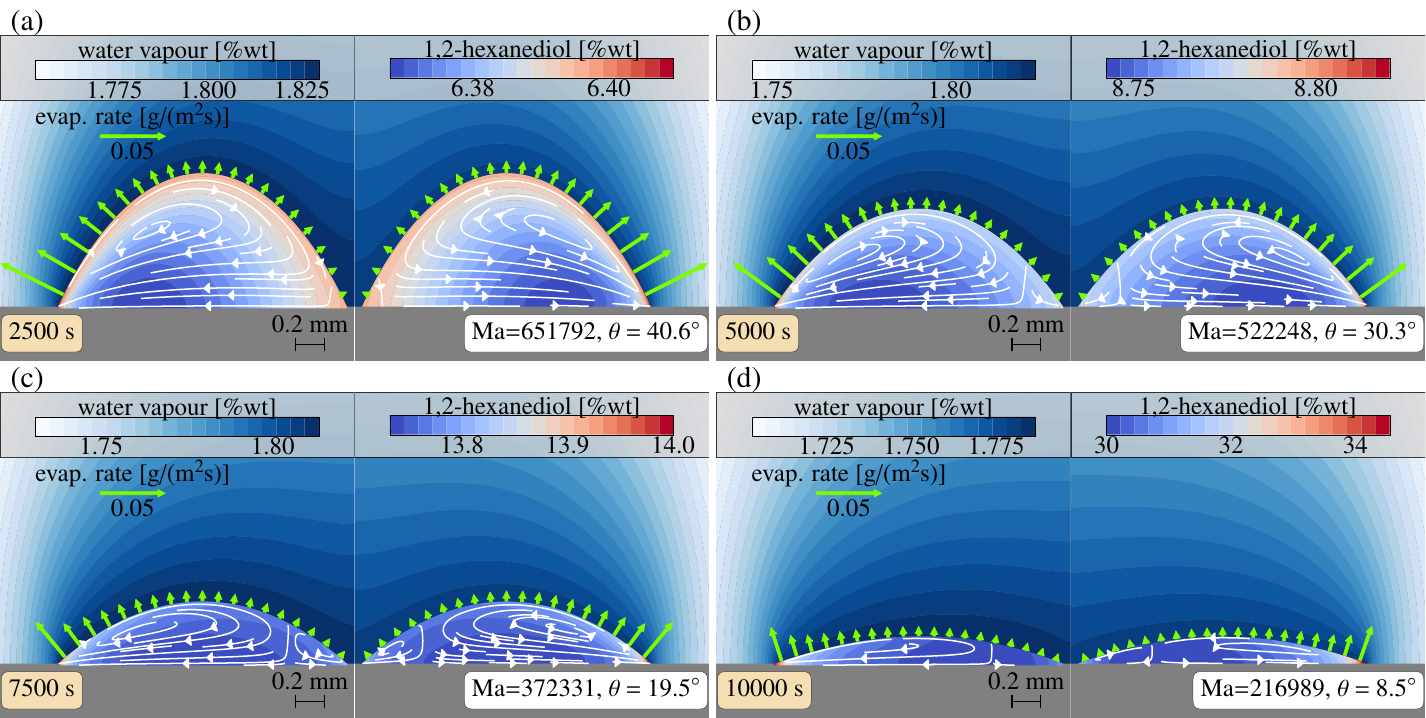}
\caption{
Perfect agreement of the quasi-stationary model (right of each subfigure) with the transient simulations (left of each subfigure) at four different times: \SI{2500}{\second} (a), \SI{5000}{\second} (b), \SI{7500}{\second} (c), and \SI{10000}{\second} (d).
Each subfigure shows the composition of the liquid phase (1,2-hexanediol mass fraction $\massfrac_\componentB$) and the velocity direction (with superimposed streamlines) in the liquid phase.
In the gas phase, the water vapour mass fraction $\cvap_\componentB$ is shown, while the liquid-gas interface highlights the evaporation flux of water.
The quasi-stationary model is evaluated at the same time as the transient simulation, using the same parameters $\MaraNumber$ and $\contactangle$.
}
\label{fig:validation_stokes}
\end{figure}

We begin by evaluating the accuracy of the quasi-stationary model (see \S\ref{sec:qs_model}) against the full transient simulations introduced in \S\ref{sec:transient_model}. 
At four distinct time instances from the transient rivulet simulations (Figure~\ref{fig:transient_rivulet}), we extract the values of $\MaraNumber$ and $\contactangle$. 
Using these parameters, we compute the corresponding quasi-stationary solutions and compare them with the transient results (see Figure~\ref{fig:validation_stokes}).

The results indicate that the quasi-stationary model replicates the key flow features very well. 
As argued by \citet{diddens2021competing} and further supported by our numerical tests, the good agreement only holds when the Marangoni number is sufficiently high. 
In contrast, at lower Marangoni numbers the ``coffee-stain'' flow becomes relevant by transporting solute to the rim, which accumulates over time as evaporation progresses, resulting in deviations from the quasi-stationary behaviour. For sufficiently large Marangoni numbers the mixing effect due to the Marangoni flow overwhelms the transport by the ``coffee-stain'' flow.

Nevertheless, even for larger Marangoni number, near the end of the drop/rivulet lifetime the ``coffee-stain'' flow will always dominate again due to a singularity in time, breaking the quasi-stationary assumption. When the contact angle approaches zero, the ``coffee-stain'' flow is forced through an increasingly small cross-sectional area, increasing the radial velocity as $1/\contactangle$. This effect is also known as ``rush hour'' \citep{marin2011}. This means that the contact angle must be sufficiently small to be able to use the lubrication model, but not too small to ensure that the dynamics remains quasi-stationary.

\begin{figure}[ht!]
\centering
\includegraphics[width=1\textwidth]{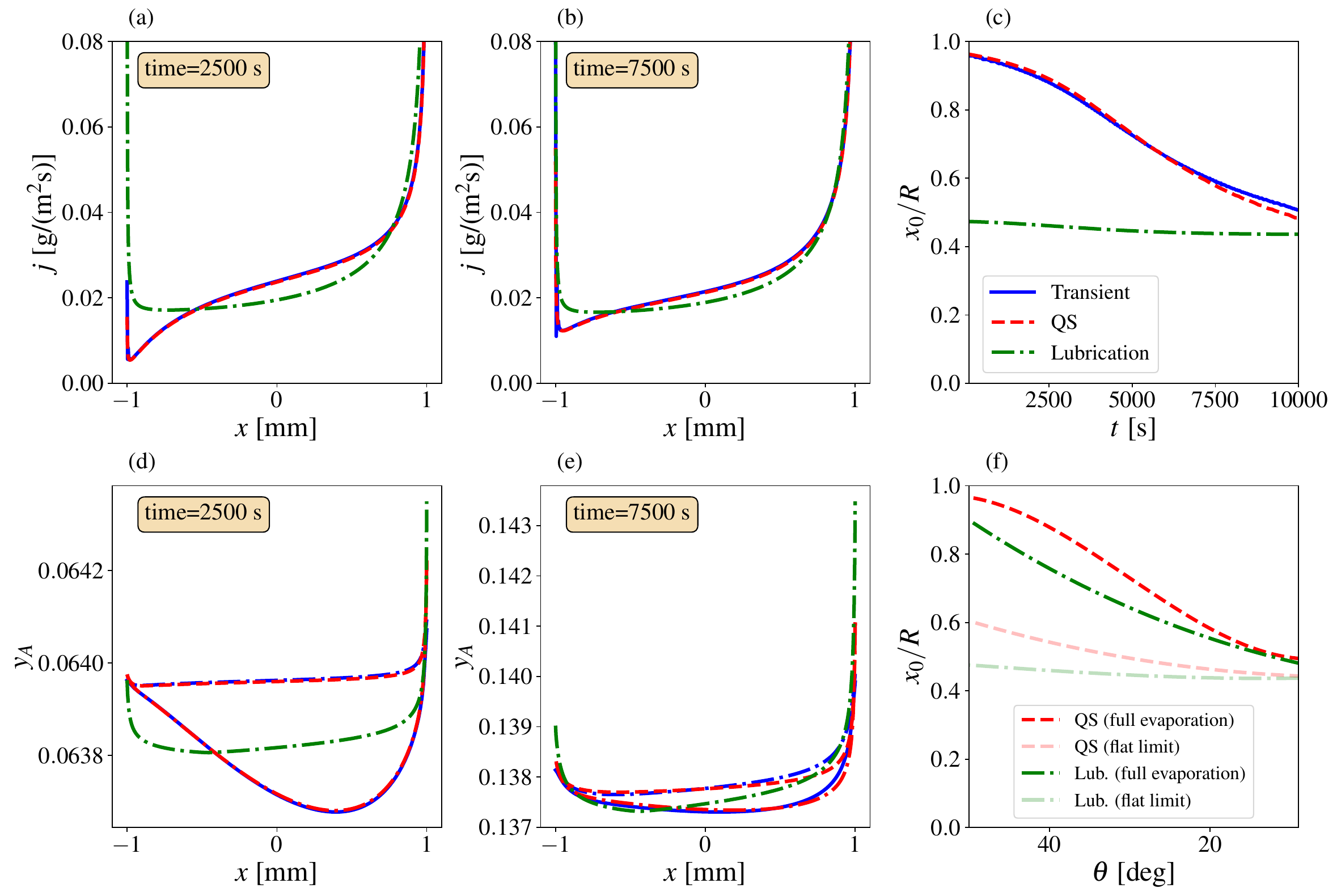}
\caption{Comparison of the lubrication model predictions (green dash-dotted) with transient (blue solid) and quasi-stationary (red dashed) simulations results: the evaporation rate at $t=$\SI{2500}{\second} (a) and $t=$\SI{7500}{\second} (b), the vertically averaged mass fraction at $t=$\SI{2500}{\second} (d) and $t=$\SI{7500}{\second} (e), the evolution of the stagnation point $\stagPoint$ over time (c). 
In (d) and (e), we show, for the transient and quasi-stationary models, the value of $\massfrac_\component$ at the substrate (lower lines) and at the liquid-gas interface (upper lines), while the lubrication model only provides the vertically averaged value.
In (f), we compare the evolution of the stagnation point $\stagPoint$ over time (or as contact angle $\contactangle$ reduces) as given by the quasi-stationary (red) and lubrication (green) models, when using expression \ref{eq:evapflux_rivulet} for the evaporation rate (transparent curves) or when using the numerically calculated evaporation rate, by solving eq.\ \ref{eq:nondim_laplace_cvap} in the gas phase for a given $\theta$ (opaque curves).
}
\label{fig:validation_lubrication}
\end{figure}

We benchmark the lubrication model for rivulets by comparing its predictions for the evaporation rate, $\evapflux$, and the vertically averaged mass fraction, $\averaged{\massfrac}_\component$, with those from both the transient and quasi-stationary models. 
Here, the both the quasi-stationary and the lubrication models use the same simulation parameters $\MaraNumber$ and $\contactangle$ extracted from the transient runs. 
Figure~\ref{fig:validation_lubrication} also presents the evolution of the stagnation point $\stagPoint$ over time (or as the contact angle $\contactangle$ reduces) as given by the quasi-stationary and lubrication models, when using expression \ref{eq:evapflux_rivulet} for the evaporation rate or when using the numerically calculated evaporation rate, by solving eq.\ \ref{eq:nondim_laplace_cvap} in the gas phase for a given $\theta$. 

For flat rivulets, the evaporation rate predicted by the lubrication model agrees well with the transient results at later times (i.e., lower contact angles), though larger discrepancies are observed at earlier times, particularly in regions near the neighbouring rivulet. 
Comparing the vertically averaged mass fraction reveals that the lubrication model's solution falls between the substrate and interfacial values from the transient and quasi-stationary models. 
Notably, for larger contact angles, the deviation of the lubrication model becomes more pronounced in the region closer to the other rivulet, likely due to differences in the evaporation flux.

The stagnation point $\stagPoint$ is not well captured by the lubrication model, particularly when using \eqref{eq:nondim_laplace_cvap} for the evaporation rate.
At earlier times (with higher contact angles), the stagnation point is significantly underestimated and remains nearly constant over time, contrasting with its evolving behaviour in both the transient and quasi-stationary models. 
At later times (with lower contact angles), the $\stagPoint$ from the transient and quasi-stationary models tends towards the value predicted by the lubrication model. 
This disagreement partially arises from the approximated evaporation flux used in the lubrication model, which does not account for the contact angle dependence. 
This is clear in figure \ref{fig:validation_lubrication}(f), where using the same evaporation rate in both the models leads to much closer agreement.
For larger contact angles, there is still some disagreement between solutions of quasi-stationary and lubrication models even when the same evaporation rate is considered, likely stemming from the assumption of negligible vertical gradients associated with lubrication model, which gets worse for larger contact angles.


\section{Shift of the stagnation point}\label{sec:shift_stagnation_point}

\subsection{Rivulets using the quasi-stationary model} \label{sec:shift_stagnation_point_rivulet}
Hereinafter, all fields are nondimensional and the tildes are dropped for brevity. Acknowledging the accuracy of the quasi-stationary model (\S\ref{sec:qs_model}) and recognising the limitations of the lubrication model (\S\ref{sec:lubrication}) for rivulets, we now investigate, with the quasi-stationary model, how the stagnation point $\stagPoint$ shifts with respect to its governing parameters, namely the contact angle $\contactangle$, the Marangoni number $\MaraNumber$, and the distance between neighbouring rivulets $b$.

\begin{figure}[ht!]
\centering
\includegraphics[width=1\textwidth]{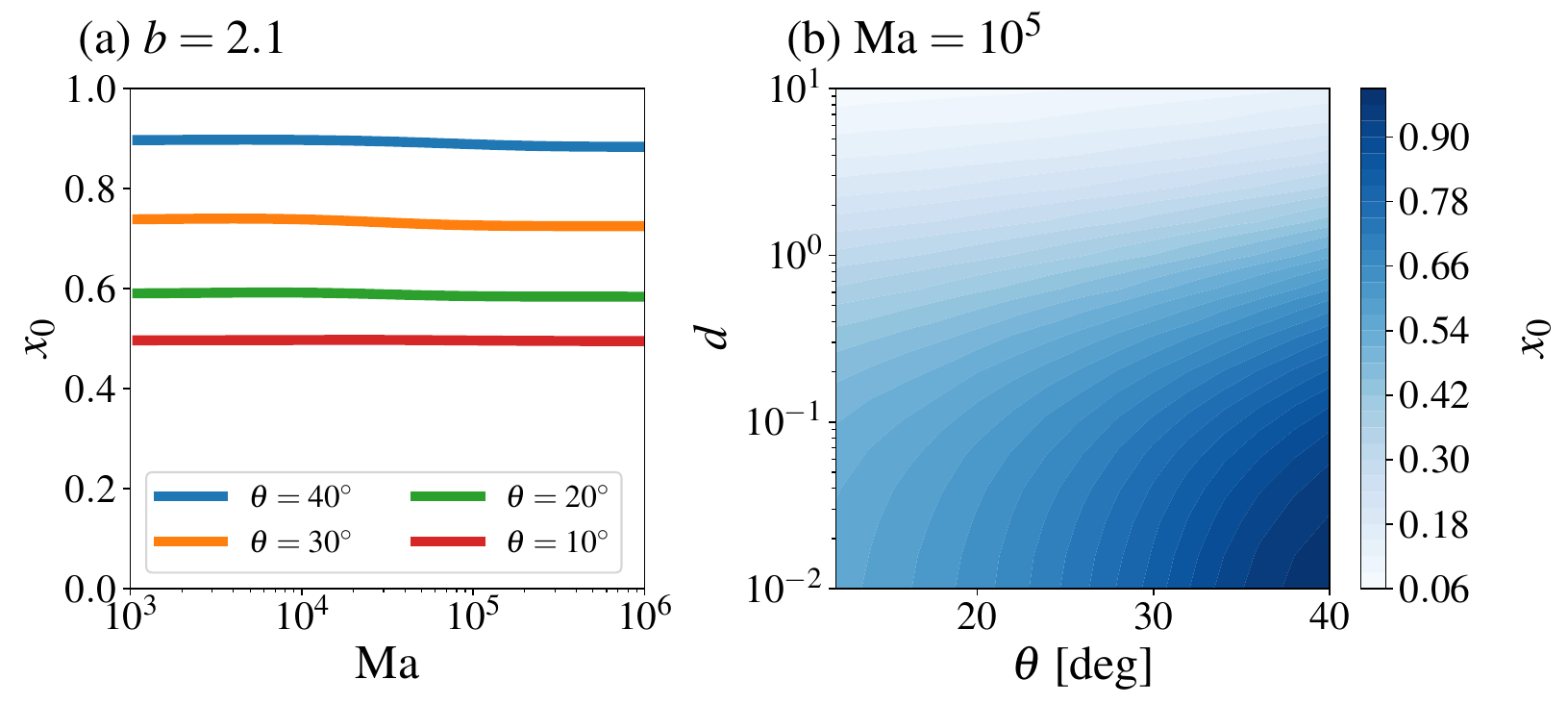}
\caption{Stagnation point in rivulets using the quasi-stationary model. 
(a) Stagnation point $\stagPoint$ versus Marangoni number for rivulets at four different contact angles: \SI{10}{\degree} (red), \SI{20}{\degree} (green), \SI{30}{\degree} (orange), and \SI{40}{\degree} (blue) and $\neighDist = 2.1$. 
(b) Phase space showing the variation of the displacement of the stagnation point $\stagPoint$ (quantifying the symmetry breaking) as a function of the contact angle $\contactangle$ and the inter-rivulet distance $\neighDist = d+2$ at fixed $\MaraNumber = 10^5$.}
\label{fig:phase_space_rivulet}
\end{figure}

Firstly, we evaluate the stagnation point $\stagPoint$ for a range of $\MaraNumber$ values at four different contact angles -- \SI{10}{\degree} (red), \SI{20}{\degree} (green), \SI{30}{\degree} (orange) and \SI{40}{\degree} (blue) -- while keeping the distance between the rivulets fixed at $\neighDist = d+2 = 2.1$, as shown in Figure~\ref{fig:phase_space_rivulet}(a). It is immediately evident that, for all considered contact angles, the stagnation point $\stagPoint$ is independent of $\MaraNumber$. An analytical explanation for this observation will be developed in subsection \S\ref{sec:analytical_rivulet}.

Secondly, we assess the stagnation point $\stagPoint$ within a phase space defined by $\contactangle$ and the distance $\neighDist$, for a fixed $\MaraNumber = 10^5$, as illustrated in Figure~\ref{fig:phase_space_rivulet}(b). 
Naturally, as $\neighDist$ increases, the behaviour of neighbouring rivulets converges towards that of an isolated rivulet, which has a stagnation point at $\stagPoint = 0$. 
Interestingly, even for a large distance $\neighDist = 10$, $\stagPoint$ is still different from zero, indicating that the neighbouring rivulet still has an influence on the flow field.
As the contact angle $\contactangle$ increases, the displacement of the stagnation point $\stagPoint$ also increases. This is expected since the shielding effect near the neighbouring rivulet is amplified with a higher contact angle since the adjacent rivulets' interfaces are closer and there is less ``space'' for the water vapour to diffuse away, i.e. the water vapour is more confined by the neighbouring surface when the contact angle increases, thereby accentuating the asymmetry in the flow for high contact angles.


\subsubsection{Analytical solutions for rivulets using the lubrication model}\label{sec:analytical_rivulet}

For rivulets we can find the shift of the stagnation point analytically when using the lubrication model, without having to solve anything numerically. To obtain an analytical solution for the $x$-shift of the stagnation point of the velocity for rivulets, we must first solve for the velocity in the drop. 
Similar as before, when using the lubrication theory, we assume that the evaporation is quasi-static and that coffee stain flow is negligible. 
Eq. \eqref{eq:nondim_stokes} in the lubrication approximation becomes 
\begin{equation}
    \frac{d p}{d x} = \frac{\partial^2 u_x}{\partial z^2}.
\end{equation}
Using the following boundary conditions, no slip at the substrate $u\big|_{z=0} = 0$, and Marangoni stress at the interface $\frac{\partial u_x}{\partial z}\big|_{z=h}=\contactangle\,\MaraNumber\frac{d\averaged{\nondimMassFrac} }{d x}$, we find the following solution
\begin{equation}
    u_x = \frac{d p}{d x}\left(z^2 - 2 zh\right) + \contactangle\,\MaraNumber\frac{d\averaged{\nondimMassFrac} }{d x}z.
\end{equation}
Using eq. \eqref{eq:nondim_flowrate} we can write the pressure gradient in terms of the total flow rate $\flowrate$ and the concentration gradient
\begin{equation}
    u_x = \contactangle\,\MaraNumber\left(\frac{3z^2}{4h} - \frac{z}{2}\right)\frac{d\averaged{\nondimMassFrac} }{d x} - \frac{3}{2}\left(\frac{z^2}{h^3} - 2\frac{z}{h^2}\right)\flowrate\bcdot \mathbf{e}_x.
\end{equation}
Due to the no-slip condition, the velocity at the substrate must be zero. Therefore, we determine the velocity close to the surface using $\frac{\partial u_x}{\partial z}$ at $z=0$ and set it to zero to find the shift of the stagnation point. When $\flowrate=0$, this condition translates to
\begin{equation}
    \frac{\partial u_x}{\partial z}\bigg|_{z=0} = \contactangle\,\MaraNumber\frac{d\averaged{\nondimMassFrac} }{d x}=0.
\end{equation}
Consequently, the stagnation point coincides with the maxima in the concentration, which we can compute starting from equation \eqref{eq:nondim_mass_fraction} with $\flowrate = 0$ and integrating over $x$
\begin{equation}
    \frac{d}{d x}\left(\Deff \height \frac{d \averaged{\nondimMassFrac}}{d x}\right) = \dfrac{\evapflux}{\contactangle} - \dfrac{\totalflux\height}{V},
\end{equation}
\begin{equation}\label{eq:dif_adv_int_1}
    \Deff \frac{d \averaged{\nondimMassFrac}}{d x} = \dfrac{1}{\contactangle\height} \int{\evapflux \,dx} - \dfrac{\totalflux}{V\height}\int \height\, dx.
\end{equation}
Knowing that $\flowrate = 0$ at the contact line, from equation \eqref{eq:nondim_film_thickness}, we find that $\flowrate = 0$ everywhere along the rivulet and therefore $\flowrate_C = -\flowrate_M$.
Equation \eqref{eq:Deff} then simplifies to
\begin{equation}
    \Deff = 1 + \contactangle^2 \dfrac{ \flowrate_M^2}{420} = 1 + \dfrac{1}{1680} h^4 \contactangle^4 \MaraNumber^2 \left( \dfrac{d \averaged{\nondimMassFrac}}{d x} \right)^2.
\end{equation}
We can then rewrite the vertically averaged mass fraction evolution equation \eqref{eq:dif_adv_int_1} as
\begin{equation}\label{eq:dif_adv_int_2}
    \frac{d \averaged{\nondimMassFrac}}{d x} +  \dfrac{1}{1680} h^4 \contactangle^4 \MaraNumber^2 \left( \dfrac{d \averaged{\nondimMassFrac}}{d x} \right)^3 = \dfrac{1}{\contactangle\height} \int{\evapflux \,dx} - \dfrac{\totalflux}{V\height}\int \height\, dx + a,
\end{equation}
with an integration constant $a$. In the limit approaching isolated rivulets ($b\gg2$), the concentration $\averaged{\nondimMassFrac}$ in the rivulet must symmetric in $x$. 
Therefore, all terms in eq. \eqref{eq:dif_adv_int_2} must be odd in $x$. 
Consequently, $a$ must be zero. 
Using the following notation: $\averaged{\nondimMassFrac}^\prime(x) = \frac{d \averaged{\nondimMassFrac}}{d x}$, $v(x) = \frac{1}{1680} \contactangle^4 \MaraNumber^2 h^4$, and $w(x)=\frac{1}{\contactangle\height} \int{\evapflux \,dx} - \frac{\totalflux}{V\height}\int \height\, dx$, we can write \eqref{eq:dif_adv_int_2} more concisely as
\begin{equation}\label{eq:dif_adv_int_3}
    \averaged{\nondimMassFrac}^\prime + v\averaged{\nondimMassFrac}^{\prime 3} = w. 
\end{equation}
Solving for $\averaged{\nondimMassFrac}^\prime$ gives 
\begin{equation}\label{eq:dif_adv_sol}
    \averaged{\nondimMassFrac}^\prime = \frac{\psi}{v} - \frac{1}{3\psi}, \quad \mathrm{with} \ \ \psi = \left[\frac{v^2 w}{2} + \left(\frac{v^3}{3^3} + \frac{v^4w^2}{2^2}\right)^\frac{1}{2}\right]^\frac{1}{3}
\end{equation}
The $x$-shift of the maxima in the concentration can be calculated by evaluating $\averaged{\nondimMassFrac}^\prime = 0$. Using \eqref{eq:dif_adv_sol} we find $v^{5/2}w=0$. Re-substituting $v$ and $w$ gives
\begin{equation}\label{eq:ana_x_shift_1}
    \left( \frac{1}{1680} \contactangle^4 \MaraNumber^2 h^4\right)^\frac{5}{2} \left( \dfrac{1}{\contactangle\height} \int{\evapflux \,dx} - \dfrac{\totalflux}{V\height}\int \height\, dx\right) = 0
\end{equation}
The first factor only depends on $x$ via the height $h$ and is positive definite everywhere except at the rim of the rivulet. Since we are interested in the non-trivial maxima/minima, we focus on the second factor, which, because of the positive-definiteness of the first factor, must be equal to zero, i.e.\
\begin{equation}\label{eq:ana_x_shift_final}
    \int{\evapflux \,dx} - \dfrac{\contactangle\totalflux}{V}\int \height\, dx= 0.
\end{equation}
Note that the above expression is independent of the Marangoni number. Even if we would have considered only Marangoni flow, by neglecting the $\averaged{\nondimMassFrac}^\prime$ term in eq. \eqref{eq:dif_adv_int_3}, or only diffusion, by neglecting the $\averaged{\nondimMassFrac}^{\prime3}$ term, we would have found the same expression for the $x$-shift. 
However, the $x$-shift cannot be expressed in terms of elementary functions without approximating of $\evapflux$ and $h$. Approximating the local evaporative flux of a neighbouring rivulet, eq. \eqref{eq:evapflux_rivulet}, for large rivulet separations $b\gg 2$ yields
\begin{equation}\label{eq:j1_order_1}
    \evapflux_{1} = \totalflux \frac{1}{\pi\sqrt{1 - x^2}}\left(1-\frac{x}{b}\right) + \mathcal{O}\left(\frac{1}{b^2}\right).
\end{equation}
Assuming that the rivulet shape is a circular arc, which is analogous to a spherical cap approximation for a drop, the rivulet shape is given by
\begin{equation}\label{eq:riv_h}
    h = \frac{1}{\contactangle}\left[\tan{\left(\frac{\contactangle}{2}\right)} - \frac{1-\sqrt{1-x^2\sin^2\contactangle}}{\sin{\contactangle}}\right].
\end{equation}
For low contact angles the rivulet height can be approximated by
\begin{equation}\label{eq:riv_h_order_3}
    h = \frac{1}{2}\left(1-x^2\right) + \frac{\contactangle^2}{24}\left(1 + 2x^2 - 3x^4\right) + \mathcal{O}\left(\contactangle^4\right).
\end{equation}
The first order term in eq. \eqref{eq:riv_h_order_3} is a parabola, whose shape is independent of the contact angle. Therefore, we need to include the next higher order term to describe the contact angle dependence of the $x$-shift. Note that this is actually a third order approximation in $\contactangle$, which is obscured by the nondimensionalisation. The volume of the rivulet is given by 
\begin{equation}\label{eq:riv_V_order_3}
V = \contactangle\int_{-1}^{1}h\,dx = \frac{2}{3}\contactangle + \frac{4}{45}\contactangle^3.
\end{equation}
Combining the first order approximation of the flux in $b^{-1}$ and the third order approximation of the rivulet height in $\contactangle$ in eq. \eqref{eq:ana_x_shift_final} and omitting the higher order terms, we find
\begin{equation}\label{eq:ana_x_shift_approx_1}
    \int{\frac{1}{\pi\sqrt{1 - x^2}}\left(1-\frac{x}{b}\right) \,dx} - \dfrac{\contactangle}{V} \int\left[\frac{1}{2}\left(1-x^2\right) + \frac{\contactangle^2}{24}\left(1 + 2x^2 - 3x^4\right)\right]\, dx= 0,
\end{equation}
\begin{equation}\label{eq:ana_x_shift_approx_2}
    \frac{\arcsin{(x)}}{\pi} - \frac{\sqrt{1-x^2}}{\pi b}-\left[\frac{x}{2} - \frac{x^3}{6} + \frac{\contactangle^2}{24}\left(x - \frac{2}{3}x^3 + \frac{3}{5}x^5\right)\right]\left(\frac{2}{3} + \frac{4}{45}\contactangle^2\right)^{-1}=0.
\end{equation}
Again, we cannot progress in terms of elementary functions without approximating further. Since we have already assumed that the rivulet separation is large, we expect that the $x$-shift will be small, allowing us to approximate eq. \eqref{eq:ana_x_shift_approx_2} for small $x$ and keep only the zeroth and first order terms in $x$,
\begin{equation}\label{eq:ana_x_shift_approx_3}
    \frac{x}{\pi} - \frac{1}{\pi b}- x \left(\frac{1}{2} + \frac{1}{24}\contactangle^2\right)\left(\frac{2}{3} + \frac{4}{45}\contactangle^2\right)^{-1}=0.
\end{equation}
Solving for $x$ finally gives the $x$-shift of the stagnation point and the maximum concentration in the neighbouring rivulet
\begin{equation}\label{eq:ana_x_shift_approx_final}
    x_0 = \frac{1}{b}\frac{15\left(16+2\contactangle^2\right)}{\left(3\pi-4\right)\left(60+5\contactangle^2\right)-12\contactangle^2}.
\end{equation}

\begin{figure}[ht!]
\centering
\includegraphics[width=0.95\textwidth]{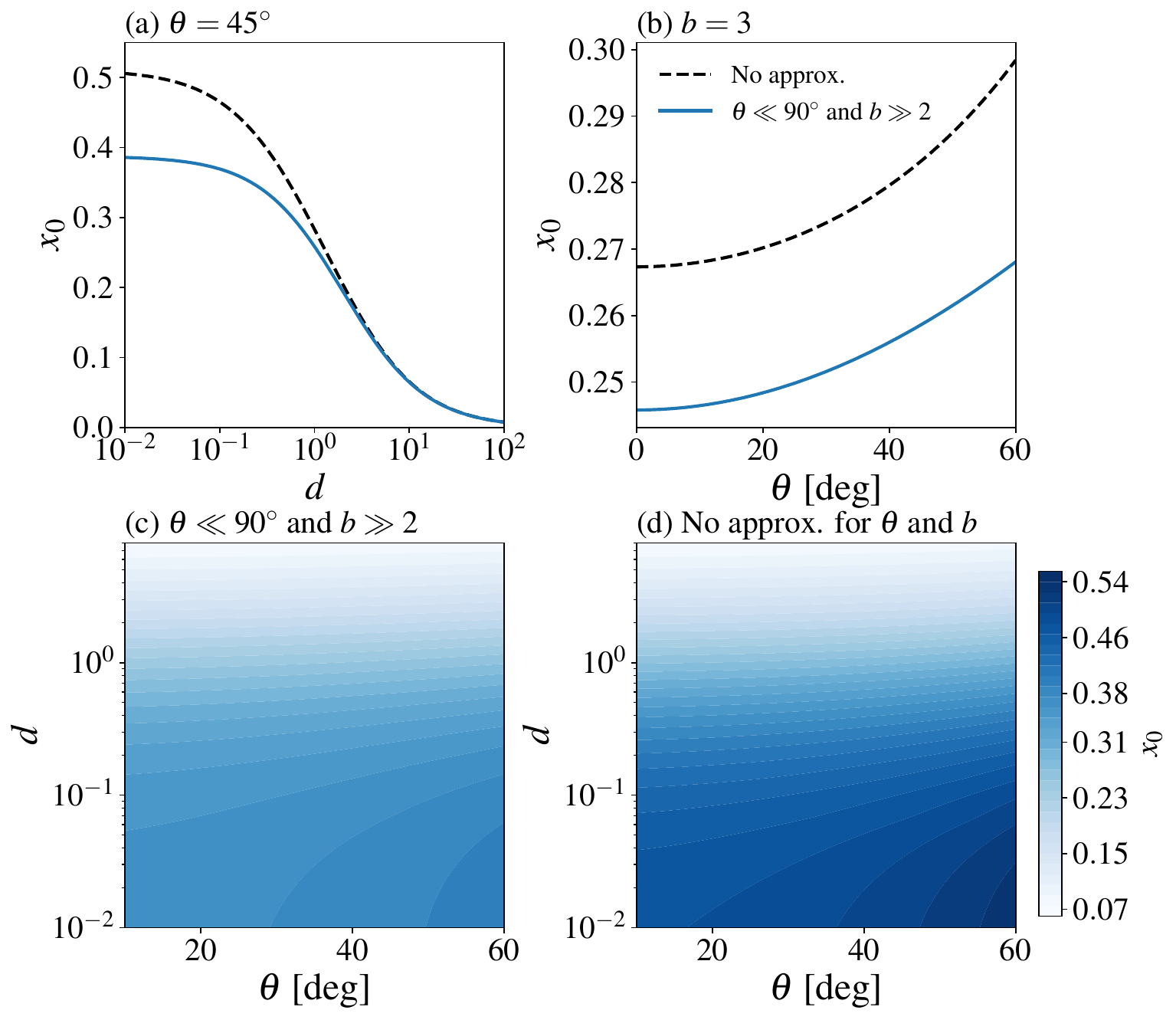}
\caption{
    Stagnation point $\stagPoint$ as a function of inter-rivulet distance $b = d + 2$ and contact angle $\contactangle$ for rivulets using the lubrication model. Panel (a) shows $\stagPoint$ for constant $\contactangle$, and panel (b) for constant $\neighDist$. The analytical approximation, eq. \eqref{eq:ana_x_shift_approx_final}, valid for small $\contactangle$ and large $\neighDist$, uses a third-order expansion for the rivulet shape and a first-order expansion in $1/\neighDist$ for the local flux. Panels (c) and (d) compare this approximation with the numerically evaluated shift obtained by solving eq. \eqref{eq:ana_x_shift_final} numerically, without simplifying the local flux or rivulet shape.
}
\label{fig:rivulets_analytical}
\end{figure}

Figure \ref{fig:rivulets_analytical} shows both the analytical approximation we just derived as well as the $x$-shift without any approximations for the local flux or rivulet shape. The latter is calculated by numerically solving eq. \eqref{eq:ana_x_shift_final} for $x$. Comparing the analytical solutions with and without approximations for small values of $d$ we find that with approximations the $x$-shift is underestimated, but the overall shape is captured accurately. The dependence on $\contactangle$ is nearly identical for both solutions, which is to be expected since we are using the lubrication approximation where we have already assumed that $\contactangle$ is small.

However, when comparing the analytical solutions of $x_0$ in figure \ref{fig:rivulets_analytical} to the numerical solutions of $x_0$ in figure \ref{fig:phase_space_rivulet}, not only do we find that the analytical solutions underestimate the numerical solutions, additionally, we also find that $x_0$ depends much more strongly on $\contactangle$. As mentioned above, the main reason for this dependency can be attributed to the evaporative flux, which should depend on the contact angle. When an analytical expression of the local flux for neighbouring rivulets with non-zero contact angles is found, it will likely be straightforward to update eq. \eqref{eq:ana_x_shift_approx_final} following the same steps as used here.


\subsection{Droplets using the lubrication model} \label{sec:shift_stagnation_point_droplet}

We employ the lubrication model to investigate the shift of the stagnation point $\stagPoint$ for droplets. 
Here, we use the contact angle dependent evaporation rate as calculated by \citet{popov2005evaporative}.
Despite the limitations of applying the lubrication model to rivulets, it has been successfully used for droplets in previous studies under the same conditions \citep{ramirez-soto_taylor_2022}.

\begin{figure}[ht!]
\centering
\includegraphics[width=1\textwidth]{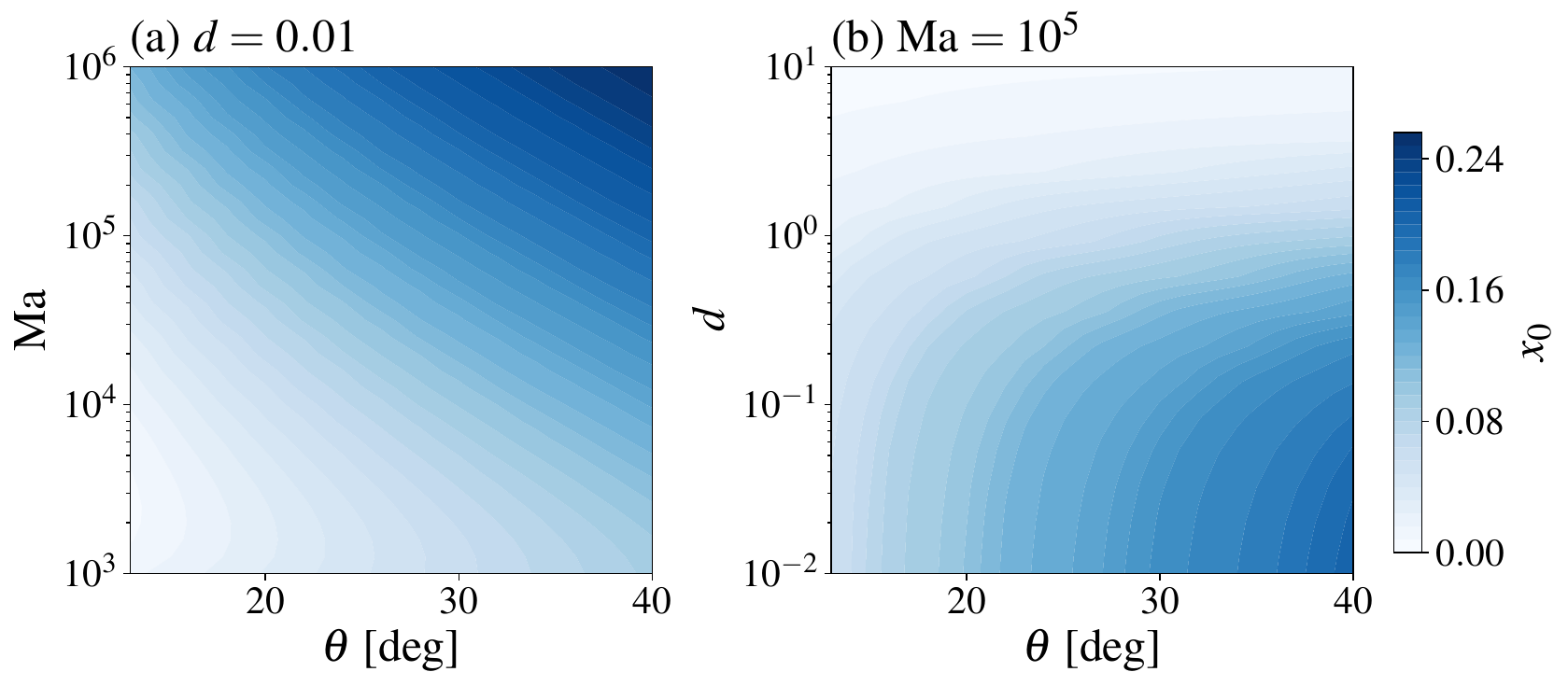}
\caption{Phase spaces showing the variation of the stagnation point $\stagPoint$ measured at the substrate as a function of: the Marangoni number ($\MaraNumber$) and contact angle ($\contactangle$) for droplets with an inter-droplet separation ($\neighDist=d+2$) of 2.01 (a); and $\contactangle$ and $\neighDist$ for droplets with fixed $\MaraNumber = 10^5$.}
\label{fig:phase_space_droplets}
\end{figure}

Notably, although the governing equations for droplets and rivulets differ only in the coordinate system of the operators, the stagnation point $\stagPoint$ is dependent on the Marangoni number $\MaraNumber$ for droplets. 
Figure~\ref{fig:phase_space_droplets} presents $\stagPoint$ in phase spaces defined by: $\MaraNumber$ and $\contactangle$ for $\neighDist=2.01$ (a); and $\neighDist$ and $\contactangle$ for $\MaraNumber = 10^5$ (b).
For sufficiently large $\contactangle$, the distance of the stagnation point $\stagPoint$ increases with $\MaraNumber$, whereas for small contact angles it displays a slightly non-monotonic trend, first decreasing and then increasing with $\MaraNumber$.

\begin{figure}[ht!]
\centering
\includegraphics[width=1\textwidth]{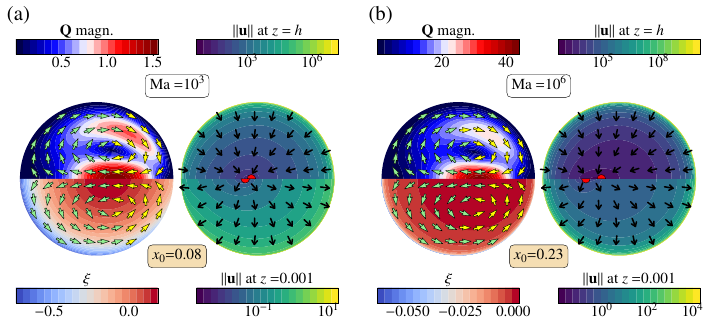}
\caption{Flow field in droplets for two different Marangoni numbers: (a) $\MaraNumber = 10^3$ and (b) $\MaraNumber = 10^6$, with a contact angle of $\contactangle = 40^\circ$ and an inter-droplet separation of $\neighDist = 2.1$. The top left panels show the flow rate $\flowrate$ (with arrows indicating its direction, coloured in yellow if dominated by Marangoni stresses, otherwise green), while the bottom left panels display the vertically averaged mass fraction $\averaged{\massfrac}_\component$. The top and bottom right panels illustrate the velocity magnitude $\|\vel\|$ at the liquid-gas interface and just above the substrate, respectively, highlighting the vertical recirculation induced by the Marangoni effect.
The stagnation point $\stagPoint$, evaluated either at the interface (top) or at the substrate (bottom), is indicated by a red dot in the right panels.}
\label{fig:flow_field_droplet}
\end{figure}

In contrast to rivulets where the Marangoni flow is largely confined to the radial or vertical directions, droplets permit Marangoni flow in the azimuthal direction as well. This additional degree of freedom leads to a more intricate evolution of the stagnation point $\stagPoint$ in droplets compared to rivulets. Figure~\ref{fig:flow_field_droplet} illustrates, for two values of $\MaraNumber$, namely $10^3$ (a) and $10^6$ (b), with a contact angle of $\contactangle = 40^\circ$ and $\neighDist = 2.1$, the volumetric flow rate $\flowrate$ (top left of each subfigure), the vertically averaged mass fraction $\averaged{\massfrac_\component}$ (bottom left), and the velocity magnitude $\|\vel\|$ as determined by the parabolic profile of the lubrication model, both at the interface (top right) and at $0.001\,R$ above the substrate (bottom right). 
In the left panels, the arrows indicate the direction of the flow rate, with the colour coding distinguishing the dominance of Marangoni-driven flow (yellow) from that of pressure-driven flow (green). 
In the right panels, the velocity directions are indicated, revealing distinct orientations near the substrate and at the interface, as expected from the vertical recirculation in droplets.

Close to the neighbouring droplet, the compositional gradients are more pronounced, resulting in a stronger Marangoni effect in that region. At the droplet axis, the flow rate $\flowrate$ is directed away from the adjacent droplet, thereby generating a backflow in the azimuthal direction.

As $\MaraNumber$ increases, the enhanced mixing within the liquid leads to a more uniform composition throughout the droplet. 
The flow rate $\flowrate$ in the backflow region becomes substantially weaker than that along the droplet's axis. 
Moreover, the position of the stagnation point $\stagPoint$ exhibits a more marked change with increasing $\MaraNumber$, particularly when evaluated at the substrate in comparison with its position at the interface, revealing large vertical gradients in the velocity. 
We attribute this shift in $\stagPoint$ to the nonlinear evolution of the compositional gradients.
However, due to the complexity of the flow field, we do not attempt to obtain a quantitative explanation for the observed behaviour of $\stagPoint$ in droplets.


\section{Conclusions} \label{sec:conclusions}
When two neighbouring binary drops evaporate, the internal flow and internal composition become asymmetric. Consequently, existing analytical and numerical models that rely on axisymmetry can no longer be used. Moreover, solving for the three-dimensional system of equations is currently computationally infeasible at a mesh resolution required to accurately resolve the evaporation rate, compositional gradients and flow dynamics. Therefore, we first investigated neighbouring cylindrical drops, i.e. rivulets, which are two-dimensional, which allowed us to test different models. By focusing on the horizontal shift of the stagnation point in the velocity close to the substrate we were able to characterise the flow and compare the different models using this single parameter.

We started by modelling the evaporation as quasi-stationary. A critical condition for this assumption to be valid is that the transport of solute via Marangoni flow must overwhelm the transport of solute via the ``coffee-stain'' flow, which would otherwise lead to the accumulation of solute over time at the rim. This translated to the requirements that the Marangoni number and the contact angle must be sufficiently large. With these conditions, perfect agreement between the transient and quasi-stationary model for rivulets in figure \ref{fig:validation_stokes} was found. Moreover, the shift of the stagnation point for rivulets only depends on the inter-rivulet distance and the contact angle and is independent of the Marangoni number as shown in figure \ref{fig:phase_space_rivulet}.

Next, we used the lubrication approximation to vertically average the system of quasi-stationary equations, while accounting for the mixing of the composition due to the Marangoni flow using the Taylor-Aris dispersion \citep{karpitschka2017marangoni}. This allowed us to analytically derive an equation for the shift of the stagnation point eq. \eqref{eq:ana_x_shift_final}, which only depends on the local flux and the drop shape and not the Marangoni number. Using a first order expansion for the local flux ($b\gg2$) and a third order expansion for the drop shape ($\contactangle\ll1$) we obtained an explicit analytical expression for the $x$-shift, eq. \eqref{eq:ana_x_shift_approx_final}.

Comparing the quasi-stationary lubrication model (figure \ref{fig:rivulets_analytical}) to the quasi-stationary Stokes model (figure \ref{fig:phase_space_rivulet}) we find that the overall agreement of the $x$-shift is qualitatively good. However, the dependence of the $x$-shift on the contact angle is not adequately captured. While part of this discrepancy comes from the lubrication approximation, which is only valid for small contact angles, most of the difference can be attributed to the local evaporative flux, which is for zero contact angle and thus does not account for any contact angle dependence. To the best of the author's knowledge, no analytical expressions of evaporating neighbouring drops/rivulets with non-zero contact angles exist in the literature and remains a topic for future work.

Finally, we applied the lubrication model to neighbouring drops. Due to the extra degree of freedom in the azimuthal direction, the total flow rate is no longer constraint, as a result the shift stagnation point now depends strongly on the Marangoni number (figure \ref{fig:phase_space_droplets}). As the Marangoni number increases the stagnation point at the substrate and at the liquid-air interface also shift relative to each other (figure \ref{fig:flow_field_droplet}). Moreover, in contrast to rivulets, and despite using the evaporative flux for flat neighbouring drops, we also get a strong dependence on the contact angle. 

We have shown that the lubrication model, which is quasi-stationary, can be used to accurately get the internal flow and internal composition of neighbouring drops/rivulets. The main limitation of this model is that ``coffee stain'' flow cannot be included as this breaks the quasi-stationary assumption. Currently, the lubrication model does not accurately describe the contact angle dependence since we only have access to the local flux for flat drops and rivulets. For neighbouring rivulets, the lateral shift of the stagnation point depends on the inter-drop distance, the contact angle, but not on the Marangoni number. 
In contrast, for neighbouring rivulets, the lateral shift does depend on the Marangoni number, owing to the extra degree of freedom in the azimuthal direction.


\section*{Funding}
This work was supported by an Industrial Partnership Programme, High Tech Systems and Materials (HTSM), of the Netherlands Organisation for Scientific Research (NWO); a funding for public-private partnerships (PPS) of the Netherlands Enterprise Agency (RVO) and the Ministry of Economic Affairs (EZ); Canon Production Printing Netherlands B.V.; University of Twente; and Eindhoven University of Technology. (project TKI HTSM - CANON - P1 - PRINTHEAD \& DROPLET FORMATION, grant no. PPS2107). 

\section*{Declaration of interests}
The authors report no conflict of interest.

\section*{Author contributions}
The authors P. J. Dekker and D. Rocha have contributed equally to this work and are both co-first author.


\bibliographystyle{jfm}
\bibliography{jfm}

\begin{thebibliography}{44}
\expandafter\ifx\csname natexlab\endcsname\relax\def\natexlab#1{#1}\fi
\def\au#1{#1} \def\ed#1{#1} \def\yr#1{#1}\def\at#1{#1}\def\jt#1{\textit{#1}} \def\bt#1{#1}\def\bvol#1{\textbf{#1}} \def\vol#1{#1} \def\pg#1{#1} \def\publ#1{#1}\def\arxiv#1{#1}\def\org#1{#1}\def\st#1{\textit{#1}}

\bibitem[Bauer {\em et~al.\/}(2002)Bauer, Frink \& Kreckel]{bauer2002introduction}
{\sc \au{Bauer, Christian}, \au{Frink, Alexander} \& \au{Kreckel, Richard}} \yr{2002}  \at{{Introduction to the GiNaC Framework for Symbolic Computation within the C++ Programming Language}}.  \jt{J. Symb. Comput.}  \bvol{33}~(1),  \pg{1--12}.

\bibitem[Brutin \& Starov(2018)]{brutin2018recent}
{\sc \au{Brutin, D} \& \au{Starov, V}} \yr{2018}  \at{Recent advances in droplet wetting and evaporation}.  \jt{Chem. Soc. Rev.}  \bvol{47}~(2),  \pg{558--585}.

\bibitem[Chong {\em et~al.\/}(2020)Chong, Li, Ng, Verzicco \& Lohse]{chong2020convection}
{\sc \au{Chong, Kai~Leong}, \au{Li, Yanshen}, \au{Ng, Chong~Shen}, \au{Verzicco, Roberto} \& \au{Lohse, Detlef}} \yr{2020}  \at{Convection-dominated dissolution for single and multiple immersed sessile droplets}.  \jt{J. Fluid Mech.}  \bvol{892},  \pg{A21}.

\bibitem[Cira {\em et~al.\/}(2015)Cira, Benusiglio \& Prakash]{cira2015vapour}
{\sc \au{Cira, Nate~J}, \au{Benusiglio, Adrien} \& \au{Prakash, Manu}} \yr{2015}  \at{Vapour-mediated sensing and motility in two-component droplets}.  \jt{Nature}  \bvol{519}~(7544),  \pg{446--450}.

\bibitem[Constantinescu \& Gmehling(2016)]{constantinescu2016further}
{\sc \au{Constantinescu, Dana} \& \au{Gmehling, J{\"u}rgen}} \yr{2016}  \at{Further development of modified {UNIFAC} ({D}ortmund): revision and extension 6}.  \jt{J. Chem. Eng. Data}  \bvol{61}~(8),  \pg{2738--2748}.

\bibitem[Deegan {\em et~al.\/}(1997)Deegan, Bakajin, Dupont, Huber, Nagel \& Witten]{deegan1997capillary}
{\sc \au{Deegan, Robert~D}, \au{Bakajin, Olgica}, \au{Dupont, Todd~F}, \au{Huber, Greb}, \au{Nagel, Sidney~R} \& \au{Witten, Thomas~A}} \yr{1997}  \at{Capillary flow as the cause of ring stains from dried liquid drops}.  \jt{Nature}  \bvol{389}~(6653),  \pg{827--829}.

\bibitem[Deegan {\em et~al.\/}(2000)Deegan, Bakajin, Dupont, Huber, Nagel \& Witten]{deegan2000contact}
{\sc \au{Deegan, Robert~D}, \au{Bakajin, Olgica}, \au{Dupont, Todd~F}, \au{Huber, Greg}, \au{Nagel, Sidney~R} \& \au{Witten, Thomas~A}} \yr{2000}  \at{Contact line deposits in an evaporating drop}.  \jt{Phys. Rev. E}  \bvol{62}~(1),  \pg{756}.

\bibitem[Dekker {\em et~al.\/}(2024)Dekker, van~der Linden \& Lohse]{dekker2024pinning}
{\sc \au{Dekker, Pim~J}, \au{van~der Linden, Marjolein~N} \& \au{Lohse, Detlef}} \yr{2024}  \at{Pinning induced motion and internal flow in neighbouring evaporating multi-component drops}.  \jt{arXiv preprint arXiv:2412.08495} .

\bibitem[Diddens {\em et~al.\/}(2024)Diddens, Dekker \& Lohse]{diddens2024non}
{\sc \au{Diddens, Christian}, \au{Dekker, Pim~J} \& \au{Lohse, Detlef}} \yr{2024}  \at{Non-monotonic surface tension leads to spontaneous symmetry breaking in a binary evaporating drop}.  \jt{arXiv preprint arXiv:2402.17452} .

\bibitem[Diddens {\em et~al.\/}(2017{\natexlab{{\em a\/}}})Diddens, Kuerten, Van~der Geld \& Wijshoff]{diddens2017modeling}
{\sc \au{Diddens, C}, \au{Kuerten, Johannes~GM}, \au{Van~der Geld, CWM} \& \au{Wijshoff, HMA}} \yr{2017{\natexlab{{\em a\/}}}}  \at{Modeling the evaporation of sessile multi-component droplets}.  \jt{J. Colloid Interface Sci.}  \bvol{487},  \pg{426--436}.

\bibitem[Diddens {\em et~al.\/}(2021)Diddens, Li \& Lohse]{diddens2021competing}
{\sc \au{Diddens, Christian}, \au{Li, Yaxing} \& \au{Lohse, Detlef}} \yr{2021}  \at{Competing {M}arangoni and {R}ayleigh convection in evaporating binary droplets}.  \jt{J. Fluid Mech.}  \bvol{914},  \pg{A23}.

\bibitem[Diddens \& Rocha(2024)]{diddens2024bifurcation}
{\sc \au{Diddens, Christian} \& \au{Rocha, Duarte}} \yr{2024}  \at{Bifurcation tracking on moving meshes and with consideration of azimuthal symmetry breaking instabilities}.  \jt{J. Comput. Phys.}  \bvol{518},  \pg{113306}.

\bibitem[Diddens {\em et~al.\/}(2017{\natexlab{{\em b\/}}})Diddens, Tan, Lv, Versluis, Kuerten, Zhang \& Lohse]{diddens2017evaporating}
{\sc \au{Diddens, Christian}, \au{Tan, Huanshu}, \au{Lv, Pengyu}, \au{Versluis, Michel}, \au{Kuerten, JGM}, \au{Zhang, Xuehua} \& \au{Lohse, Detlef}} \yr{2017{\natexlab{{\em b\/}}}}  \at{Evaporating pure, binary and ternary droplets: thermal effects and axial symmetry breaking}.  \jt{J. Fluid Mech.}  \bvol{823},  \pg{470--497}.

\bibitem[Eggers \& Pismen(2010)]{eggers2010nonlocal}
{\sc \au{Eggers, Jens} \& \au{Pismen, Len~M}} \yr{2010}  \at{Nonlocal description of evaporating drops}.  \jt{Phys. Fluids}  \bvol{22}~(11).

\bibitem[Fabrikant(1985)]{fabrikant1985potential}
{\sc \au{Fabrikant, VI}} \yr{1985}  \at{On the potential flow through membranes}.  \jt{Zeitschrift f{\"u}r angewandte Mathematik und Physik ZAMP}  \bvol{36}~(4),  \pg{616--623}.

\bibitem[Gelderblom {\em et~al.\/}(2022)Gelderblom, Diddens \& Marin]{gelderblom2022evaporation}
{\sc \au{Gelderblom, Hanneke}, \au{Diddens, Christian} \& \au{Marin, Alvaro}} \yr{2022}  \at{Evaporation-driven liquid flow in sessile droplets}.  \jt{Soft Matter}  \bvol{18}~(45),  \pg{8535--8553}.

\bibitem[Hatte {\em et~al.\/}(2019)Hatte, Pandey, Pandey, Chakraborty \& Basu]{hatte2019universal}
{\sc \au{Hatte, Sandeep}, \au{Pandey, Keshav}, \au{Pandey, Khushboo}, \au{Chakraborty, Suman} \& \au{Basu, Saptarshi}} \yr{2019}  \at{Universal evaporation dynamics of ordered arrays of sessile droplets}.  \jt{J. Fluid Mech.}  \bvol{866},  \pg{61--81}.

\bibitem[Heil \& Hazel(2006)]{heil2006oomph}
{\sc \au{Heil, Matthias} \& \au{Hazel, Andrew~L}} \yr{2006} {oomph-lib--An Object-Oriented Multi-Physics Finite-Element Library}.  \bt{In {\em Fluid-structure interaction: Modelling, simulation, optimisation\/}},  \pg{pp. 19--49}. Springer.

\bibitem[Hu \& Larson(2005)]{hu2005analysis}
{\sc \au{Hu, Hua} \& \au{Larson, Ronald~G}} \yr{2005}  \at{Analysis of the effects of {M}arangoni stresses on the microflow in an evaporating sessile droplet}.  \jt{Langmuir}  \bvol{21}~(9),  \pg{3972--3980}.

\bibitem[Hu \& Larson(2006)]{hu2006marangoni}
{\sc \au{Hu, Hua} \& \au{Larson, Ronald~G}} \yr{2006}  \at{{M}arangoni effect reverses coffee-ring depositions}.  \jt{J. Phys. Chem. B}  \bvol{110}~(14),  \pg{7090--7094}.

\bibitem[Karpitschka {\em et~al.\/}(2017)Karpitschka, Liebig \& Riegler]{karpitschka2017marangoni}
{\sc \au{Karpitschka, Stefan}, \au{Liebig, Ferenc} \& \au{Riegler, Hans}} \yr{2017}  \at{Marangoni contraction of evaporating sessile droplets of binary mixtures}.  \jt{Langmuir}  \bvol{33}~(19),  \pg{4682--4687}.

\bibitem[Khilifi {\em et~al.\/}(2019)Khilifi, Foudhil, Fahem, Harmand \& Ben]{khilifi2019study}
{\sc \au{Khilifi, Dorra}, \au{Foudhil, Walid}, \au{Fahem, Kamel}, \au{Harmand, Souad} \& \au{Ben, Jabrallah}} \yr{2019}  \at{Study of the phenomenon of the interaction between sessile drops during evaporation}.  \jt{Therm. Sci.}  \bvol{23}~(2 Part B),  \pg{1105--1114}.

\bibitem[Kim \& Stone(2018)]{kim2018direct}
{\sc \au{Kim, Hyoungsoo} \& \au{Stone, Howard~A}} \yr{2018}  \at{Direct measurement of selective evaporation of binary mixture droplets by dissolving materials}.  \jt{J. Fluid Mech.}  \bvol{850},  \pg{769--783}.

\bibitem[Laghezza {\em et~al.\/}(2016)Laghezza, Dietrich, Yeomans, Ledesma-Aguilar, Kooij, Zandvliet \& Lohse]{laghezza2016collective}
{\sc \au{Laghezza, Gianluca}, \au{Dietrich, Erik}, \au{Yeomans, Julia~M}, \au{Ledesma-Aguilar, Rodrigo}, \au{Kooij, E~Stefan}, \au{Zandvliet, Harold~JW} \& \au{Lohse, Detlef}} \yr{2016}  \at{Collective and convective effects compete in patterns of dissolving surface droplets}.  \jt{Soft matter}  \bvol{12}~(26),  \pg{5787--5796}.

\bibitem[Lohse(2022)]{lohse2022fundamental}
{\sc \au{Lohse, Detlef}} \yr{2022}  \at{Fundamental fluid dynamics challenges in inkjet printing}.  \jt{Annu. Rev. Fluid Mech.}  \bvol{54}~(1),  \pg{349--382}.

\bibitem[Lohse \& Zhang(2020)]{lohse2020physicochemical}
{\sc \au{Lohse, Detlef} \& \au{Zhang, Xuehua}} \yr{2020}  \at{Physicochemical hydrodynamics of droplets out of equilibrium}.  \jt{Nat. Rev. Phys.}  \bvol{2}~(8),  \pg{426--443}.

\bibitem[Mampallil \& Eral(2018)]{mampallil2018review}
{\sc \au{Mampallil, Dileep} \& \au{Eral, Huseyin~Burak}} \yr{2018}  \at{A review on suppression and utilization of the coffee-ring effect}.  \jt{Adv. Colloid Interface Sci.}  \bvol{252},  \pg{38--54}.

\bibitem[Marin {\em et~al.\/}(2019)Marin, Karpitschka, Noguera-Mar{\'\i}n, Cabrerizo-V{\'\i}lchez, Rossi, K{\"a}hler \& Rodr{\'\i}guez~Valverde]{marin2019solutal}
{\sc \au{Marin, Alvaro}, \au{Karpitschka, Stefan}, \au{Noguera-Mar{\'\i}n, Diego}, \au{Cabrerizo-V{\'\i}lchez, Miguel~A}, \au{Rossi, Massimiliano}, \au{K{\"a}hler, Christian~J} \& \au{Rodr{\'\i}guez~Valverde, Miguel~A}} \yr{2019}  \at{Solutal {M}arangoni flow as the cause of ring stains from drying salty colloidal drops}.  \jt{Phys. Rev. Fluids}  \bvol{4}~(4),  \pg{041601}.

\bibitem[Mar\'{\i}n {\em et~al.\/}(2011)Mar\'{\i}n, Gelderblom, Lohse \& Snoeijer]{marin2011}
{\sc \au{Mar\'{\i}n, \'Alvaro~G.}, \au{Gelderblom, Hanneke}, \au{Lohse, Detlef} \& \au{Snoeijer, Jacco~H.}} \yr{2011}  \at{Order-to-disorder transition in ring-shaped colloidal stains}.  \jt{Phys. Rev. Lett.}  \bvol{107},  \pg{085502}.

\bibitem[Masoud \& Felske(2009)]{masoud2009analytical}
{\sc \au{Masoud, Hassan} \& \au{Felske, James~D}} \yr{2009}  \at{Analytical solution for {S}tokes flow inside an evaporating sessile drop: {S}pherical and cylindrical cap shapes}.  \jt{Phys. Fluids}  \bvol{21}~(4).

\bibitem[Masoud {\em et~al.\/}(2021)Masoud, Howell \& Stone]{masoud2021evaporation}
{\sc \au{Masoud, Hassan}, \au{Howell, Peter~D} \& \au{Stone, Howard~A}} \yr{2021}  \at{Evaporation of multiple droplets}.  \jt{J. Fluid Mech.}  \bvol{927},  \pg{R4}.

\bibitem[Pahlavan {\em et~al.\/}(2021)Pahlavan, Yang, Bain \& Stone]{pahlavan2021evaporation}
{\sc \au{Pahlavan, Amir~A}, \au{Yang, Lisong}, \au{Bain, Colin~D} \& \au{Stone, Howard~A}} \yr{2021}  \at{Evaporation of binary-mixture liquid droplets: the formation of picoliter pancakelike shapes}.  \jt{Phys. Rev. Lett.}  \bvol{127}~(2),  \pg{024501}.

\bibitem[Picknett \& Bexon(1977)]{picknett1977evaporation}
{\sc \au{Picknett, RG} \& \au{Bexon, R}} \yr{1977}  \at{The evaporation of sessile or pendant drops in still air}.  \jt{J. Colloid Interface Sci.}  \bvol{61}~(2),  \pg{336--350}.

\bibitem[Popov(2005)]{popov2005evaporative}
{\sc \au{Popov, Yuri~O.}} \yr{2005}  \at{Evaporative deposition patterns: Spatial dimensions of the deposit}.  \jt{Phys. Rev. E}  \bvol{71},  \pg{036313}.

\bibitem[Ramírez-Soto \& Karpitschka(2022)]{ramirez-soto_taylor_2022}
{\sc \au{Ramírez-Soto, O.} \& \au{Karpitschka, S.}} \yr{2022}  \at{Taylor dispersion in thin liquid films of volatile mixtures: {A} quantitative model for {Marangoni} contraction}.  \jt{Phys. Rev. Fluids}  \bvol{7}~(2),  \pg{L022001}.

\bibitem[Rocha {\em et~al.\/}(2024)Rocha, Lohse \& Diddens]{rocha2024marangoni}
{\sc \au{Rocha, Duarte}, \au{Lohse, Detlef} \& \au{Diddens, Christian}} \yr{2024}  \at{Marangoni flow driven hysteresis and azimuthal symmetry breaking in evaporating binary droplets}.  \jt{arXiv preprint arXiv:2412.17096} .

\bibitem[Schofield {\em et~al.\/}(2020)Schofield, Wray, Pritchard \& Wilson]{schofield2020shielding}
{\sc \au{Schofield, Feargus~GH}, \au{Wray, Alexander~W}, \au{Pritchard, David} \& \au{Wilson, Stephen~K}} \yr{2020}  \at{The shielding effect extends the lifetimes of two-dimensional sessile droplets}.  \jt{J. Eng. Math.}  \bvol{120}~(1),  \pg{89--110}.

\bibitem[Scriven \& Sternling(1960)]{scriven1960marangoni}
{\sc \au{Scriven, LE} \& \au{Sternling, CV}} \yr{1960}  \at{The marangoni effects}.  \jt{Nature}  \bvol{187}~(4733),  \pg{186--188}.

\bibitem[Sefiane(2014)]{sefiane2014patterns}
{\sc \au{Sefiane, Khellil}} \yr{2014}  \at{Patterns from drying drops}.  \jt{Adv. Colloid Interface Sci.}  \bvol{206},  \pg{372--381}.

\bibitem[Sneddon(1966)]{sneddon1966mixed}
{\sc \au{Sneddon, Ian~Naismith}} \yr{1966} {\em Mixed boundary value problems in potential theory\/}.  \publ{North-Holland, Amsterdam}.

\bibitem[Wilson \& D'Ambrosio(2023)]{wilson2023evaporation}
{\sc \au{Wilson, Stephen~K} \& \au{D'Ambrosio, Hannah-May}} \yr{2023}  \at{Evaporation of sessile droplets}.  \jt{Annu. Rev. Fluid Mech.}  \bvol{55}~(1),  \pg{481--509}.

\bibitem[Wray {\em et~al.\/}(2020)Wray, Duffy \& Wilson]{wray2020competitive}
{\sc \au{Wray, Alexander~W}, \au{Duffy, Brian~R} \& \au{Wilson, Stephen~K}} \yr{2020}  \at{Competitive evaporation of multiple sessile droplets}.  \jt{J. Fluid Mech.}  \bvol{884},  \pg{A45}.

\bibitem[Yarin {\em et~al.\/}(2006)Yarin, Szczech, Megaridis, Zhang \& Gamota]{yarin2006lines}
{\sc \au{Yarin, AL}, \au{Szczech, JB}, \au{Megaridis, CM}, \au{Zhang, J} \& \au{Gamota, DR}} \yr{2006}  \at{Lines of dense nanoparticle colloidal suspensions evaporating on a flat surface: Formation of non-uniform dried deposits}.  \jt{J. Colloid Interface Sci.}  \bvol{294}~(2),  \pg{343--354}.

\bibitem[Zang {\em et~al.\/}(2019)Zang, Tarafdar, Tarasevich, Choudhury \& Dutta]{zang2019evaporation}
{\sc \au{Zang, Duyang}, \au{Tarafdar, Sujata}, \au{Tarasevich, Yuri~Yu}, \au{Choudhury, Moutushi~Dutta} \& \au{Dutta, Tapati}} \yr{2019}  \at{Evaporation of a droplet: From physics to applications}.  \jt{Phys. Rep.}  \bvol{804},  \pg{1--56}.

\end{thebibliography}

\end{document}